\documentclass[preprint,12pt]{elsarticle}

\usepackage{graphicx}
\usepackage{subfigure}
\usepackage{float}
\usepackage{enumitem}
\usepackage{amsmath}

\usepackage{amssymb}

\begin{document}

\begin{frontmatter}



\title{ The coupling effect between the environment and strategies drives the emergence of group cooperation}

\author[inst1]{Changyan Di}
\affiliation[inst1]{organization={School of Information Science and Engineering,Lanzhou University}      }    
           
\author[inst1]{Qingguo Zhou}

\author[inst2]{Jun Shen}
\affiliation[inst2]{organization={School of Computing and Information Technology,University of Wollongong}}

\author[inst1]{Jinqiang Wang}

\author[inst1]{Rui Zhou}

\author[inst1]{Tianyi Wang}

\begin{abstract}
Introducing environmental feedback into evolutionary game theory has led to the development of eco-evolutionary games, which have gained popularity due to their ability to capture the intricate interplay between the environment and decision-making processes. However, current researches in this field focus on the study to macroscopic evolutionary dynamics in infinite populations. In this study, we propose a multi-agent computational model based on reinforcement learning to explore the coupled dynamics between strategies and the environment in finite populations from a bottom-up perspective. Our findings indicate that even in environments that favor defectors, high levels of group cooperation can emerge from self-interested individuals, highlighting the significant role of the coupling effect between the environment and strategies. Over time, the higher payoff of defection can be diluted due to environmental degradation, while cooperation can become the dominant strategy when positively reinforced by the environment. Remarkably, individuals can accurately detect the inflection point of the environment solely through rewards, when a reinforcing positive feedback loop are triggered, resulting in a rapid increase in agents' rewards and facilitating the establishment and maintenance of group cooperation. Our research provides a fresh perspective on understanding the emergence of group cooperation and sheds light on the underlying mechanisms involving individuals and the environment.
\end{abstract}

\begin{graphicalabstract}
\includegraphics{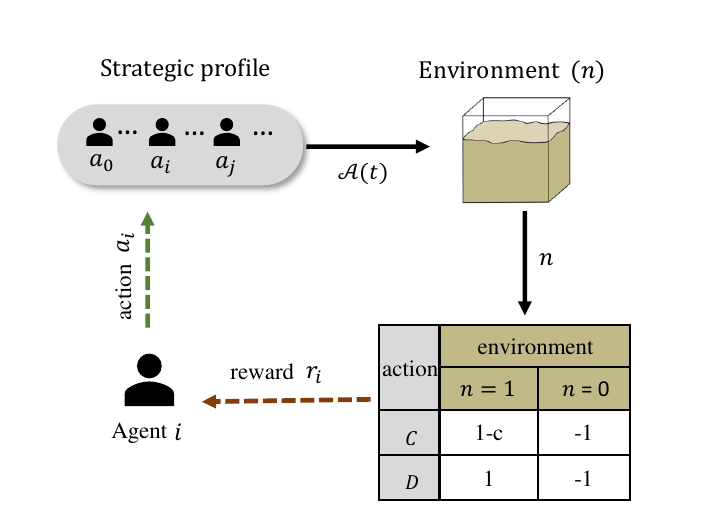}
\end{graphicalabstract}

\begin{highlights}
\item The coupling effect between macro environment and individual behavior is the key factor to solve the social dilemma. In a static environment, rewards of different strategies are compared simultaneously, leading to a social dilemma due to the higher payoff of defection compared to cooperation. However, when individuals are placed in a dynamic environment that is coupled with their actions, we find that the expected payoffs of different strategies are not fixed but undergo dynamic changes. The higher expected payoff of defection can be diluted over time due to environmental degradation caused by an excessive number of defectors, while cooperation may become the dominant strategy if positively reinforced by environmental feedback.

\item Group cooperation emerges as a direct result of a mutually reinforcing positive feedback loop among the environment, immediate rewards, and individual actions (or group states). Despite the agents' lack of awareness regarding the macro-level context, they possess the ability to astutely discern the inflection point of the environment solely through their rewards. This pivotal moment prompts agents to experience a surge in immediate rewards, thereby triggering a positive feedback loop among the environment, their rewards, and their current actions. Consequently, cooperation emerges within the group.

\end{highlights}

\begin{keyword}
Emergence of cooperation\sep social dilemma\sep reinforcement learning \sep eco-evolutionary game
\PACS 02.50.Le - Decision theory and game theory
\PACS 87.23.Ge - Dynamics of social systems
\PACS 87.23.Cc - Population dynamics and ecological pattern formation

\end{keyword}

\end{frontmatter}


\section{Introduction}

In the tragedy of the commons (TOC)\cite{hardin1968tragedy}, individuals are driven to deplete resources in pursuit of their own self-interest, leading to a collective decline. This inherent conflict between the interests of the collective and those of the individual, commonly known as the social dilemma, is a central challenge in the field of cooperative evolution, and is a recurring theme in various game-theoretic models such as the Prisoner's Dilemma\cite{sachs2004evolution, lacey2008prisoners} and the Public Goods Game\cite{santos2008social, hauert2008public}. As cheaters who reap the benefits of cooperation without bearing the costs gain a competitive advantage, each individual is tempted to betray the common good in pursuit of their own gain\cite{doebeli2005models}, guided by the principle of ``survival of the fittest". Paradoxically, cooperation is a ubiquitous phenomenon observed in natural systems, ranging from simple microorganisms to complex mammals\cite{greig2004prisoner, rainey2003evolution, mitri2011social, maclean2006resource, turner1999prisoner}, from social insects to human societies. 

In the powerful framework of evolutionary game theory \cite{nowak2004emergence}, various mechanisms have been proposed to explain the emergence of cooperation\cite{nowak2006five,taylor2007transforming} over the past few decades, including direct reciprocity, indirect reciprocity\cite{leimar2001evolution,nowak2005evolution}, kin selection\cite{smith1964group}, network reciprocity\cite{nowak1992evolutionary,may2006network,hauert2004spatial} and group selection\cite{williams2018adaptation}. Some other studies attribute the reasons for cooperation to subjective altruism or prosociality\cite{bicchieri2010behaving,henrich2010markets,hughes2018inequity}   when human society are taken as a reference. However, if the tragedy can only be avoided when higher-level incentives are invoked, why non-human organisms can avoid overexploiting the resources on which they depend\cite{falster2003plant,wenseleers2004tragedy,foster2004diminishing}.

In recent years, the novel concept of environmental feedback has been proposed to explain the emergence and maintenance of cooperation\cite{weitz2016oscillating,szolnoki2018environmental,lin2019spatial}. As it captures the complex interplay between cooperative behavior and the environment, there has been a growing interest in incorporating environmental feedback into evolutionary game theory\cite{hauert2006evolutionary,huang2015stochastic}. In particular, in certain scenarios, the co-evolution of players' strategies and their environment can give rise to oscillating dynamics between cooperators and defectors\cite{weitz2016oscillating}. However, these raise two issues. Firstly, if cooperators consistently face a disadvantage in each environment, the effectiveness of environmental feedback in preventing the extinction of cooperators may be diminished. Secondly, previous researches focus on the evolution of collective behavior at the macroscopic level in an infinite population. That lacks a microscopic perspective to comprehensively compare and analyze the emergence of cooperation, particularly in the context of group cooperation such as joint hunting and joint defense, at the level of individual behaviors.

\begin{figure*}[!t]
    \centering
    \includegraphics[width=0.8\linewidth]{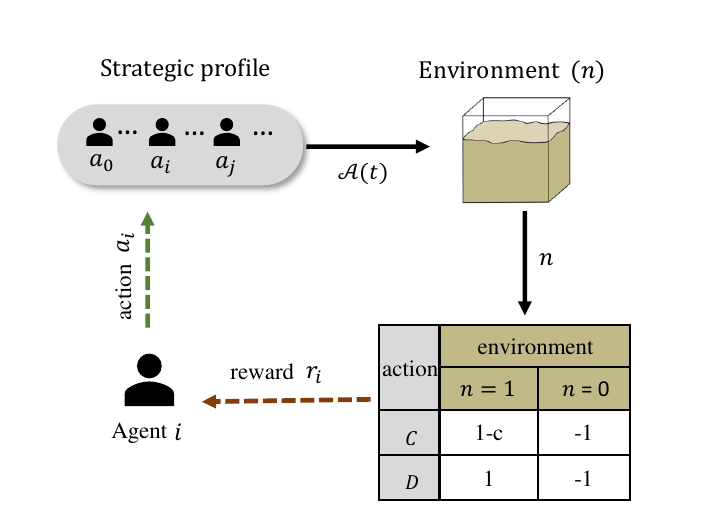}
    \caption{Schamatic of the coupled relations between individual actions and the macro environment. GCM is a multi-agent system consisting of $N$ identical agents. At each time step $t$, agent $i$ selects an action from her strategy set $\mathcal{S}$, which includes cooperation (referred to as $C$) and defection (referred to as $D$). The strategic profile $\mathcal{A}(t)$ of the population comprises all agents' actions, denoted as $\mathcal{A}(t) = \left \{a_0, a_2, ..., a_i, ..., a_{N-1} \right \}$. The macro environment ($n$) is determined or influenced by $\mathcal{A}(t)$ and can be described by functions $n=f(\mathcal{A}(t), n)$. Depending on the specific environmental states, agent $i$ receives varying rewards. Each agent continuously learns from their past rewards $r_i(t)$ and aims to maximize their future payoff. However, regardless of the conditions, the payoff of strategy $D$ is never lower than that of strategy $C$, illustrating the presence of a social dilemma. The parameter $c$ represents the cost of cooperation relative to defection.}
    \label{fig:model}
\end{figure*}

Therefore, we constructed a multi-agent computational model shown in Fig.\ref{fig:model}, inspired by animal cognitive behavioral experiments, to investigate the coupled dynamics between individual decision-making and the environment. There are two reasons for our multi-agent computational model: 1) Computational models provide a concise and clear way to describe the coupling scenarios between strategies and environment, providing a complete description of the dynamic process of individual decision-making and group evolution. 2) Researches on coupled dynamics predominantly relies on replicator equations\cite{cardillo2010co,nagashima2019stochastic}, which compare the expected payoffs of different decision-making groups at a macroscopic level, often resulting in the disappearance of cooperators. In contrast, agent-based models adopt a different paradigm, envisioning a world in which decision-making is decentralized. That allows for individual agents to be observed as objects of study, facilitating comparative analysis of the emergence and evolution of cooperation from both macro(group) and micro(individual) perspectives. 

In our investigation, we discovered that the coupling effect between the macro environment and individual behavior is the key factor in resolving social dilemmas. In an interactive environment, the higher expected payoff of defection can be gradually diluted over time due to environmental degradation caused by an excessive number of defectors. Meanwhile, cooperation may emerge as the dominant strategy when positively reinforced by environmental feedback. From an individual's perspective, cooperation is a rational outcome of self-interest. Furthermore, individuals with empirical learning ability exhibit remarkable discernment in recognizing the inflection point of the environment. When this pivotal moment occurs, a mutually reinforcing positive feedback loop is established among the environment, the agent's rewards, and her actions. That is the direct reason for the formation of group cooperation.

\section{The model}

The group coupling model(called GCM) is a multi-agent system with $N$ identical agents, each of whom is endowed with reinforcement learning to explore the intrinsic relations between rewards and actions. The ultimate aim of agents are trying their best to achieve a higher possible payoff. The basic designs are explained below:

\subsection{The payoff matrix of GCM}

Unlike classical games where the payoffs associated with strategies remain fixed, the eco-evolutionary game introduces a dynamic framework. Within this framework, the rewards of different actions vary across distinct environments ($n$), as illustrated in Fig.\ref{fig:model}. At each time step $t$, every agent selects an action $a_i(t)$, which collectively forms the strategic profile of the group, denoted as $\mathcal{A}(t)$. This profile influences, and in some cases, determines the state of the macro environment ($n$), which is then fed back to the agent through her returns $r_{a_i}(t)$. Since an agent's rewards depend on the strategies of other individuals, GCM is game-theoretic naturally, with the payoff matrix serving as a crucial factor for understanding the game model.

Despite the diversity of fields, the social dilemma shares the common feature that the payoff of cooperators get lower than that of defectors, and selfishness reduces the resource over which individuals are competing and lowers group fitness. As shown in Fig.\ref{fig:model}, the payoff matrix signifies that every agent thrives in a rich environment, reaping a reward of $+1$. Nonetheless, cooperators bear an expense indicated by the parameter $c$, relative to defectors, in order to include the essence of the social dilemma. Conversely, in a poor environment, all agents endure adversity and are met with a detrimental reward of $-1$. This pattern of payoffs occurs across scales from microbes to humans in the PGG and commons' dilemmas\cite{levin2014public,weitz2016oscillating}. For example, overfishing of marine resources can lead to the depletion of fishery resources, which affects the income of the entire industry. In contrast, when fishery resources are abundant, cheaters can gain more by abusing the resources than those who follow the norms of reasonable fishing\cite{tilman2017maintaining}.

\subsection{The learning method of agents}
The GCM is a homogeneous system in which each agent has the same strategy set $\mathcal{S}$, learning method, and decision-making process. At each step $t$, agents take actions $a_i(t)$ and receive a corresponding reward $r_{a_i}(t)$. There are two steps for agents to determine their actions for the next step $t+1$:

\begin{enumerate}
    \item Firstly, each agent updates her expected payoff value, denoted as $\pi_{a_i}$, that is : 
    \begin{equation}\label{learning}
        \pi_{a_i}(t+1) \longleftarrow \pi_{a_i}(t) + l \cdot \left[r_{a_i}(t) - \pi_{a_i}(t)\right]
    \end{equation}

    where $i$ is the index of the agent within the group and $l$ is the learning step size \cite{tuyls2003selection}. The parameter $l$ allows to adjust the proportion of importance between current and historical rewards in the learning process, allowing each agent to exhibit adaptive behavior based on past experience.
    
    It is important to distinguish between strategies, which refer to the combination of optional actions for an agent, and actions, which represent the actual choices made by the agent.

    \item   Secondly, each agent uses the $\epsilon$-greedy algorithm to guiding their actions \cite{geng2022reinforcement}, which is a probabilistic method that balances exploitation and exploration. That is,

    \begin{equation}\label{greedy}
    \begin{aligned}
     a_i(t+1)= \begin{cases}\underset{a}{\operatorname{argmax}}\left(\pi_C, \pi_D\right), \text { with } & p=1-\epsilon, \\ \,C \text { or } \, D \quad \text { randomly, with } & p=\epsilon,\end{cases}
    \end{aligned}
    \end{equation}

    Here, $\epsilon$ is the exploration parameter, typically ranging from $1\%$ to $5\%$. Under a high probability of $1-\epsilon$, each agent chooses the action with the maximum expected payoff, demonstrating a pursuit of reward maximization. However, there is still a small probability $\epsilon$ that an agent will choose a random action to explore the possibility of higher returns.

\end{enumerate}

Typically, the predominant update rules used in game model research include the proportional imitation rule, the Moran-like rule (also called the death-birth or birth-death rule), the Fermi rule, and others \cite{cardillo2010co,nagashima2019stochastic}. They compare the payoffs of different strategies at the same time, where strategy $D$ is clearly superior to strategy $C$. With the recent rapid advances in artificial intelligence (AI), reinforcement learning algorithms have emerged as novel update methods under investigation in the field of evolutionary games \cite{zhang2020oscillatory,geng2022reinforcement}. These algorithms allow agents to achieve maximum rewards through trial and error. Although there are so many studies using Q-learning algorithm in game models, they considered different update object\cite{macy2002learning} and state sets\cite{geng2022reinforcement}.

In contrast, our focus is not on the internal state of the agent, but only on the expected payoffs of different strategies. In this respect, agents only need to consider information about actions and their corresponding payoffs. As a result, the learning method allows agents to make decisions based on empirical induction, even in the absence of global knowledge. Given a sufficient amount of data about actions and payoffs, strategies that yield higher long-term payoffs are reinforced over time and eventually become dominant. Conversely, strategies associated with higher immediate rewards may be diluted and eventually replaced by dominant strategies.

\subsection{The environment}
In the GCM, the environment $n$ is determined or influenced by the strategic profile of the group $\mathcal{A}(t)$, that is

\begin{equation}\label{env}
    n(t+1)= f(\mathcal{A}(t),n(t))  
\end{equation}
The state of the environment $n(t)$ lies between $0$ and $1$. The higher $n(t)$ indicates a richer environment. Thus, by manipulating this function, we can simulate different environments and explore the dynamic decision-making processes of agents in response to varying environmental conditions. In our paper, we distinguish between two different environmental functions. 

\subsubsection{State-dependent environment}

    In a state-dependent environment, the environment is determined by the cooperative level of the population $p_C$, which is derived from $\mathcal{A}(t)$. Inspired by the group deer hunting game, the relational function of $p_C$ and $n$ can be expressed as follows:
    
    \begin{equation}\label{SDE}
        n=\left\{\begin{matrix}1\quad \text{if } p_C\ge T
      & \\0\quad \text{if } p_C< T,\,
      &
    \end{matrix}\right.
    \end{equation}
    
     where $T$ is the threshold. Eq.\ref{SDE} clarifies that if the proportion of cooperators in the group exceeds the threshold $T$, the collective goal of the group, i.e.successful hunting, can be achieved, which indicates the formation of group cooperation.  Conversely, when the number of cooperators falls below the threshold that prevents the collective goal from being achieved, the environment becomes impoverished and individuals gain $-1$. Under such circumstances, the environment exhibits a discrete binary state of either $0$ or $1$, with no intermediate state.

\subsubsection{Resource-dependent environment}
 
In the resource-dependent system, the cooperative level in the population does not directly determine the state of the environment, but rather influences its improvement or degradation. A commonly used model to capture the evolution of that kind is formulated as follows\cite{weitz2016oscillating,tilman2020evolutionary}:

\begin{equation}\label{fx}
     \dot{n}(t) =k_1*n(t)*[1-n(t)]*g(p_C) ,
\end{equation}

 where $g(p_C)$ elucidates the strategy-dependent feedback mechanism of the environment, and the sign of $g(p_C)$ indicates whether $n(t)$ decreases or increases, corresponding to environmental degradation or enhancement. In this study, we assume $g(p_C)=\theta *p_C-p_D=(1+\theta)p_C-1$\cite{weitz2016oscillating}, in which $\theta > 0$ is the ratio of the enhancement rates to degradation rates of cooperators and defectors. Finally, the rate of environmental dynamics is determined by the dimensionless quantity $k_1$, such that when $0< k_{1} \ll 1$, the environmental change is relatively gradual compared to the frequency of strategy changes\cite{weitz2016oscillating}. In such cases, the environmental state is modeled as a continuously changing value.

This function can be used to simulate the phenomenon described in TOC, in which the environment may represent a natural resource, such as the concentration of a critical nutrient, fish stocks, or grazing land, among others. Regardless of the specific natural resource in question, the general principle is that the environment tends to improve with a greater presence of cooperators and, conversely, to deteriorate with more non-cooperators.

\section{Results}
We observe persistent oscillations of the group cooperation and the environmental state in GCM under both the state-dependent environment and resource-dependent environment. Remarkably, our model is a homogeneous system with no spatial structure, where agents pursue their self-interest to the fullest. Even if the agent is only aware of her own rewards, group cooperation can still emerge under social dilemma, which is very different from the results by replicator equations. In that case, resource degradation is inevitable, i.e., $n$ converges to 0, which is also the evolutionary stability equilibrium(ESS) of the payoff matrix in the GCM(Fig.\ref{si:phase_plane}\subref{pp1}). Further, we find this kind of oscillation exists for a relatively wide range of parameters(Fig.\ref{fig:parameter}). Following we hightlight the key factors for the emergence of group cooperation .

\begin{figure*}[!t]
    \centering
     \subfigure[Oscillatory dynamics of GCM under the state-dependent environment]{\includegraphics[width=\textwidth]{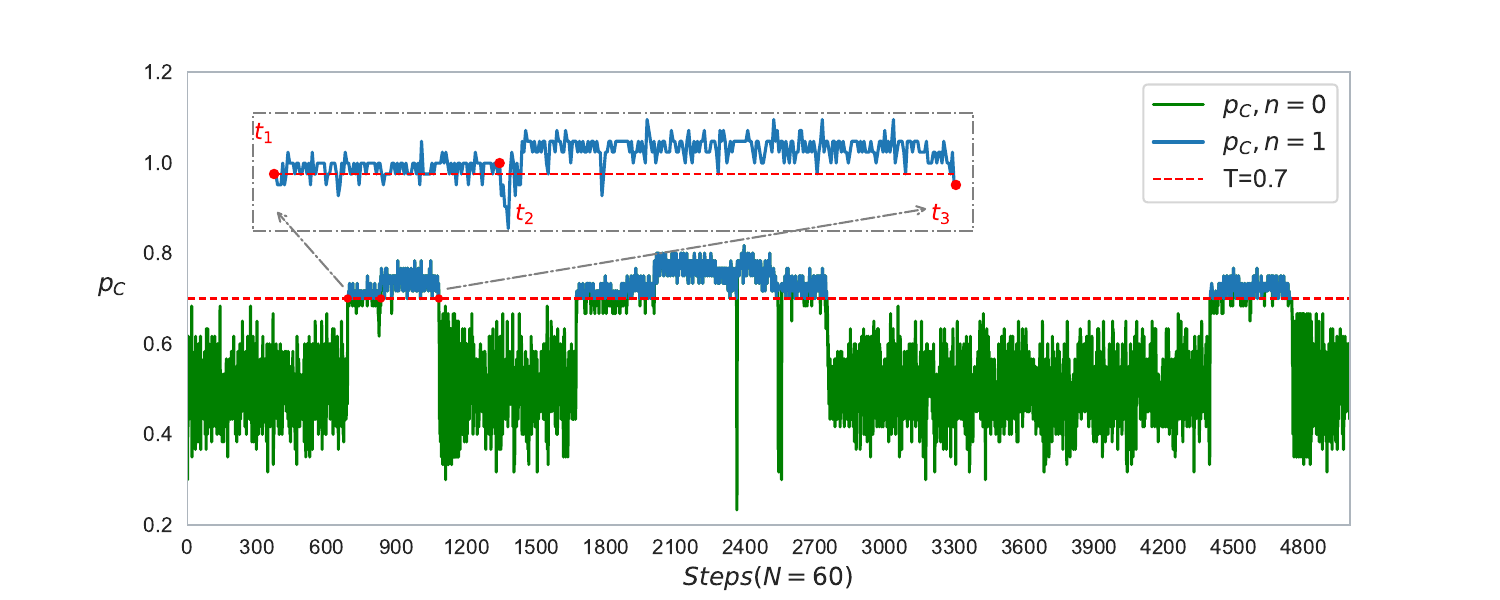}\label{Oscillatory}} 
  
    \subfigure[Cellular graph of the group between $t=630\sim 710$]{\includegraphics[width=0.45\textwidth]{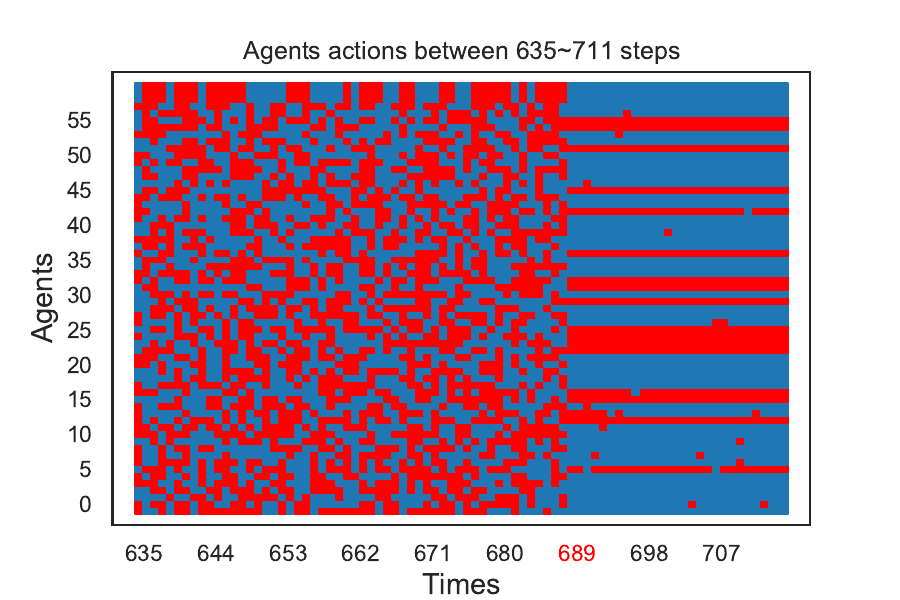}\label{cell1 in SDE}} 
    \subfigure[Cellular graph of the group between $t=1030\sim 1110$]{\includegraphics[width=0.45\textwidth]{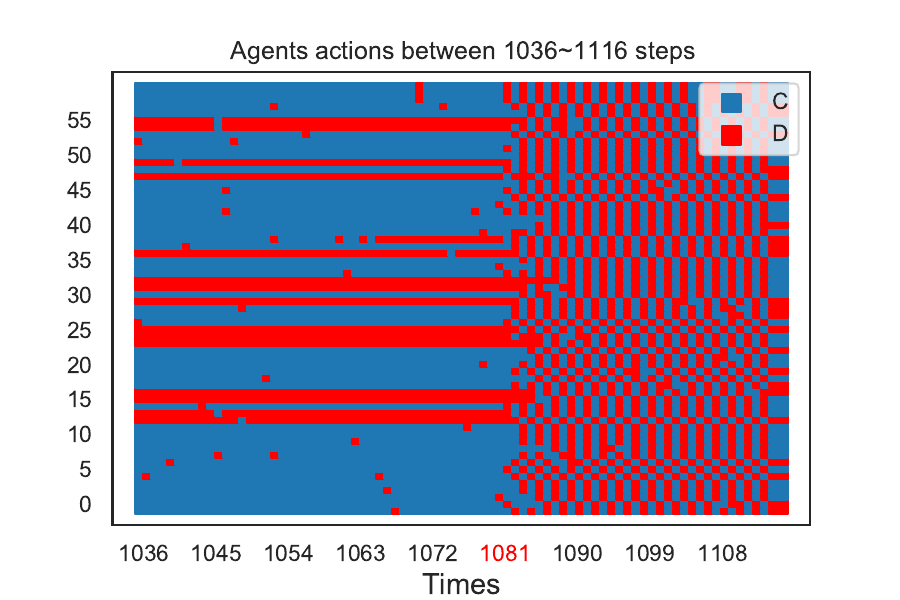}\label{cell2 in SDE}}
       
    \caption{Persistent oscillations of the fraction of cooperators from the macro perspective with parameters $N=60, T=0.7, \epsilon=0.03, l=0.2,c=0.2$. (a) Time series of the proportion of cooperators, denoted $p_C$, in the state-dependent environment. The threshold is represented by the red dashed line. Notably, three key time points ($t_1, t_2, t_3$) mark the formation, sharp shock, and collapse of group cooperation. (b)(c) The cellular diagram shows the emergence of order from a dynamically changing state of disorder at $t_1=689$, followed by a collapse at $t_3=1081$. The discrete red and blue dots on the graph denote the exploratory actions of the agents.}
    \label{fig:Oscillatory under SDE}
 \end{figure*}

 \begin{figure*}[!tph]
    \centering
    
    \subfigure[$\pi$ dynamic of Agent10]{\includegraphics[width=0.48\textwidth]{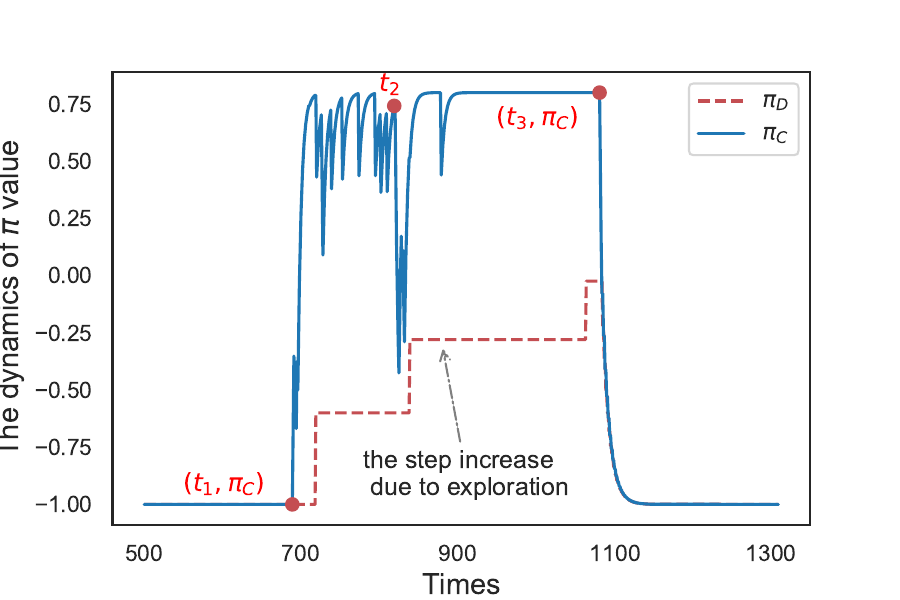}\label{q14}} 
    \subfigure[$\pi$ dynamic of Agent6]{\includegraphics[width=0.48\textwidth]{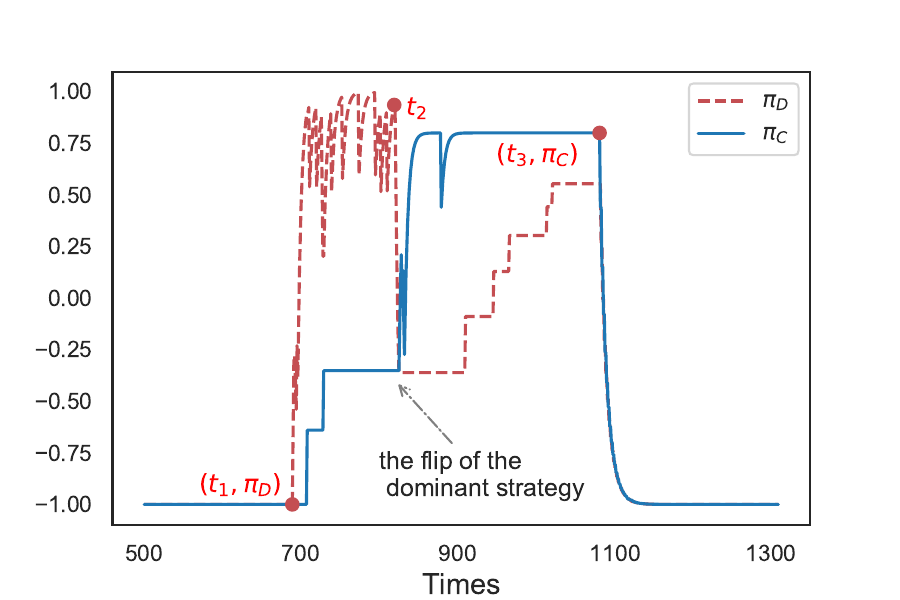}\label{q59}}
       
    \caption{$\pi$ dynamics from the micro perspective. Despite using the same learning and decision algorithm, the agents show different dynamics in their $\pi$ values and actions. Agent10 predominantly adopts strategy $C$ (indicated by the blue curve) from time $t_1$, but occasionally adopts exploratory action $D$, leading to a  step increase in the red dot curve after receiving a reward of $+1$. However, Agent6 initially adopts strategy $D$ at the time $t_1$, but switches to strategy $C$ at $t=825$ due to the disruption observed during $t=820\sim 824$.}
    \label{fig:dynamic q under SDE}
 \end{figure*}
 
\subsection{Oscillation of the system state under the coupled condition}

We use the state-dependent environment as an example to elucidate the underlying causes of the emergence and collapse of group cooperation. In fact, both environments share the same law regarding the emergence of group cooperation. We will describe one cycle in detail, examining it from both macro-group and micro-individual perspectives.

\subsubsection{The formation of group cooperation} 

It is intriguing to observe that at time $t_1=689$ the environment $n$ undergoes an abrupt shift from $0$ to $1$ due to $p_C>T$, as intuitively depicted in Figure \ref{fig:Oscillatory under SDE}\subref{Oscillatory}, signifying a sudden transformation of the group from disorder to order(Fig.\ref{fig:Oscillatory under SDE}\subref{cell1 in SDE}). To explain this drastic change, we examined the behavior of the agents and the group leading up to this pivotal moment.

During the interval denoted by $T_A=0\sim 688$, the level of group cooperation $p_C$ remains below the threshold $T$, resulting in all agents receiving a payoff of $-1$, regardless of their actions. As the game iterates, $\pi_{C,i}$ and $\pi_{D,i}$, which denote the expected payoffs of different strategies of agent $i$, rapidly approach $-1$, prompting agents to resort to randomized actions using the $\epsilon\-$ greedy algorithm. At the macro level, the group, which consists of $N$ agents that can choose actions between cooperation ($C$, denoted by $0$) and defection ($D$, denoted by $1$), contains $2^N$ different microstates, ranging from $\underbrace{000 \ldots 0}_{N}$ to $\underbrace{111 \ldots 1}_{N}$. The behavior of the group exhibits a stochastic exploration in the state space of $N$ agents, which is a typical binomial distribution $X \sim B(N,p)$. When $N*p>10$, the distribution can be approximated by a normal distribution with $\mu=30$ and $\sigma^2=15$, consistent with the statistical distribution diagram during this interval (Fig.\ref{fig:normal_sde}).
     
 At the time $t_1=689$, the collective state coincidentally reaches the threshold condition, causing an abrupt change in the environmental state. This change is quickly communicated to the agents by a positive reward. As a result, the actions taken by the agents at this point become dominant shown in Fig.\ref{fig:dynamic q under SDE} at $t_1$ and the group state metamorphoses from chaos to order(Fig.\ref{fig:Oscillatory under SDE}\subref{cell1 in SDE}).

From a macro perspective, the group as a whole persistently explores different states by trial and error, seeking to optimize rewards while interacting with the macro environment. When favorable feedback is received, the group state stabilizes accordingly. Intuitively, the group state supported by the environment emerges triumphantly from serendipitous exploration.

\begin{figure*}[tp!]
    \centering
    \subfigure[The distribution of $\Delta\pi_i(t=820)$]{\includegraphics[width=0.4\linewidth]{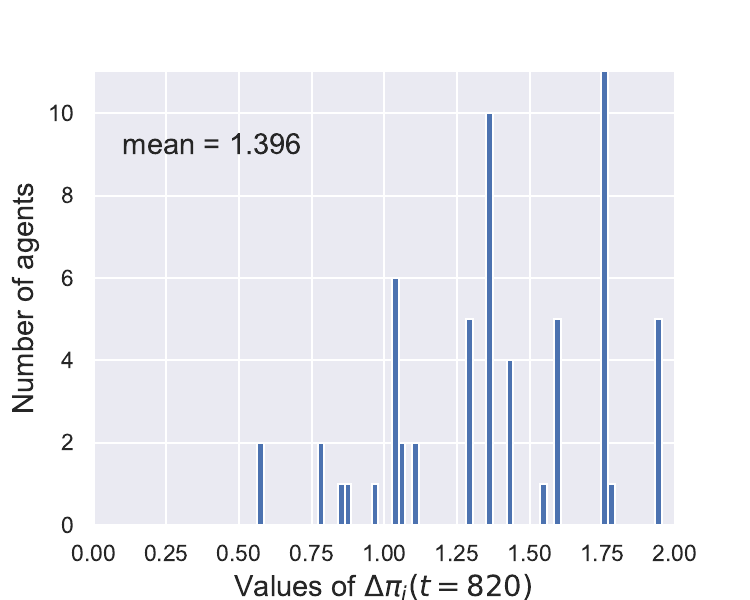}\label{hist820}} 
    \subfigure[The distribution of $\Delta\pi_i(t=1080)$ ]{\includegraphics[width=0.4\linewidth]{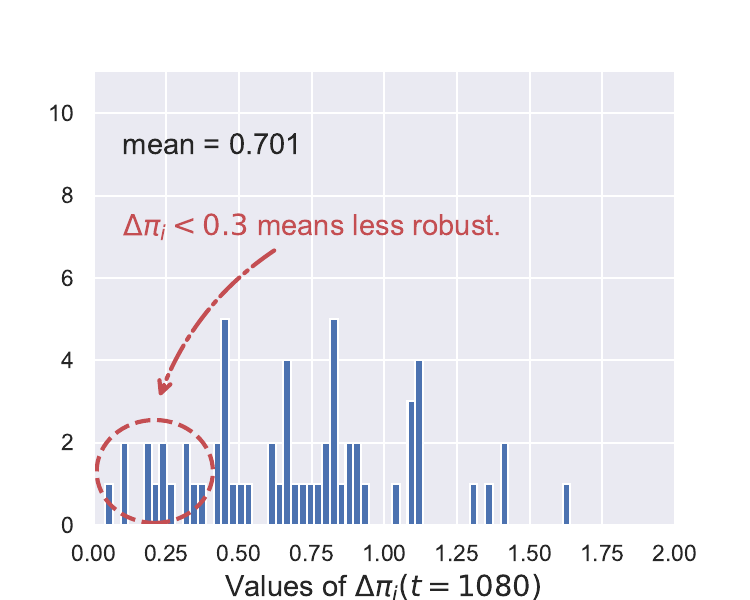}\label{hist1080}}
    
    \caption{Comparisons between the distributions of $\Delta \pi_i(t=820)$ and $\Delta \pi_i(t=1080)$ indicate different anti-interference capabilities. (a) The distribution of $\Delta \pi_i(t=820)$ shows a distinct concentration in the range of $1$ to $2$, with a minimum value not falling below $0.5$. (b) $\Delta \pi_i(t=1080)$ is distributed over the interval $0$ to $1.25$, and in particular the $\Delta \pi_i$ values of these agents (indexed as 59, 27, 46, 40, etc.) are below $0.3$, which suggests that even a small perturbation has the potential to completely disrupt the ordered state of the group.}
    
    \label{fig:distribution}
 \end{figure*}

\subsubsection{The collapse of group cooperation}

$t_3=1081$ marks the critical point beyond which the collective state falls into complete disorder. Intriguingly, in Fig.\ref{fig:Oscillatory under SDE}\subref{Oscillatory}, the cooperation level consistently exceeds the threshold before this critical time, implying remarkable stability in group cooperation. That is, there was no warning at the macro level before the collapse of cooperation. However, a closer examination of the $\pi$ dynamics at the individual level reveals something different.

Given that agents have a very low probability to choose an exploratory strategy rather than the dominant one during any time, that is about $N*\epsilon=1.8$ agents on average. These instances manifest as discrete red and blue dots in Fig.\ref{fig:Oscillatory under SDE}\subref{cell1 in SDE}\subref{cell2 in SDE}, occurring between $t_1=689$ and $t_3=1081$. Notably, if these actions do not alter the environmental state, they are positively rewarded, leading to incremental increases in the $\pi$ value of the dominated strategies for agents, as shown in Fig.\ref{fig:dynamic q under SDE}\subref{q14}\subref{q59}. That is a gradual convergence between $\pi_{C,i}$ and $\pi_{D,i}$ across all agents. Let us define
\begin{equation}
\begin{aligned}
\Delta \pi_i\left(t\right)=\begin{vmatrix} \pi_{C,i}\left(t\right)-\pi_{D,i}\left(t\right)
\end{vmatrix},
\end{aligned}
\end{equation}
as an indicator of agent $i$'s resistance to perturbations. A higher value of $\Delta \pi_i\left(t\right)$ signifies a more stable performance of the agent, and a lower probability of changing its dominant strategy in the face of temporary chaos.
 
 To illustrate the sudden collapse of system state, we compare the distribution of $\Delta \pi_i$ for the group at two critical times, $t=820$ and $t=1080$. The group's cooperation recovers quickly, rising to a level of $0.73$ after a brief period of turbulence at $t=820$, while it races toward complete disorder at $t=1080$. As shown in Fig. \ref{fig:distribution}, the distribution of $\Delta \pi_i\left(t=820\right)$ for the group is concentrated between $1$ and $2$, with a minimum value not less than $0.5$. This indicates that when a larger number of agents stochastically adopt exploratory strategies simultaneously, resulting in environmental changes and subsequent chaos, the group can quickly recover through effective fine-tuning.
 
 In comparison, the distribution of $\Delta \pi_i\left(t=1080\right)$ is observed to range between $0$ and $1.25$, with a mean of $0.701$, which is significantly lower than that of $t=820$. Notably, there are up to $9$ agents whose $\Delta \pi_i$ is less than $0.3$, indicating that even small perturbations can lead to a complete breakdown of group cooperation. It is observed in Fig.\ref{fig:Oscillatory under SDE}\subref{cell2 in SDE} that those agents with $\Delta \pi_i<0.3$ first reversed their actions around $t=1080$, leading to the complete collapse of the ordered state of the group.

We can use an analogy to illustrate the breakdown vividly. In contrast to the overt actions $a_i(t)$ of agent $i$, we perceive the fluctuations in the value of $\pi_{C,i}$ or $\pi_{D,i}$ as her underlying mental activities. In other words, from the moment when group cooperation is formed, the seeds of betrayal have been sown and germinated. Until the values of $\Delta \pi$ for a few agents approach zero, the aggregation effect occurs, which ultimately leads to the collapse of the system state.

After $t_3=1081$, the expected payoff of either strategy $C$ or strategy $D$ experiences a sharp decline, causing the group to revert to a state of random exploration similar to that observed during interval $T_A$. Consequently, the group's performance undergoes a repetitive pattern of oscillation.

In the above analysis, we have thoroughly investigated the underlying determinants that contribute to the emergence and dissolution of group cooperation in a state-dependent environment. The stochastic behaviors of the agents are stabilized by the positive feedback from the environment, ultimately leading to the adoption of these behaviors as dominant strategies. As a result, the formerly turbulent dynamics give way to coherent group states. However, it may give us the illusion that the emergence of group cooperation is caused by abrupt changes in the environment. In the following, we will analyze their relationship between them in detail in a continuous environment.

 \begin{figure*}[!tph]
    \centering 
    
    \subfigure[Oscillations of the system state under the resource-dependent environment]{\includegraphics[width=0.9\textwidth]{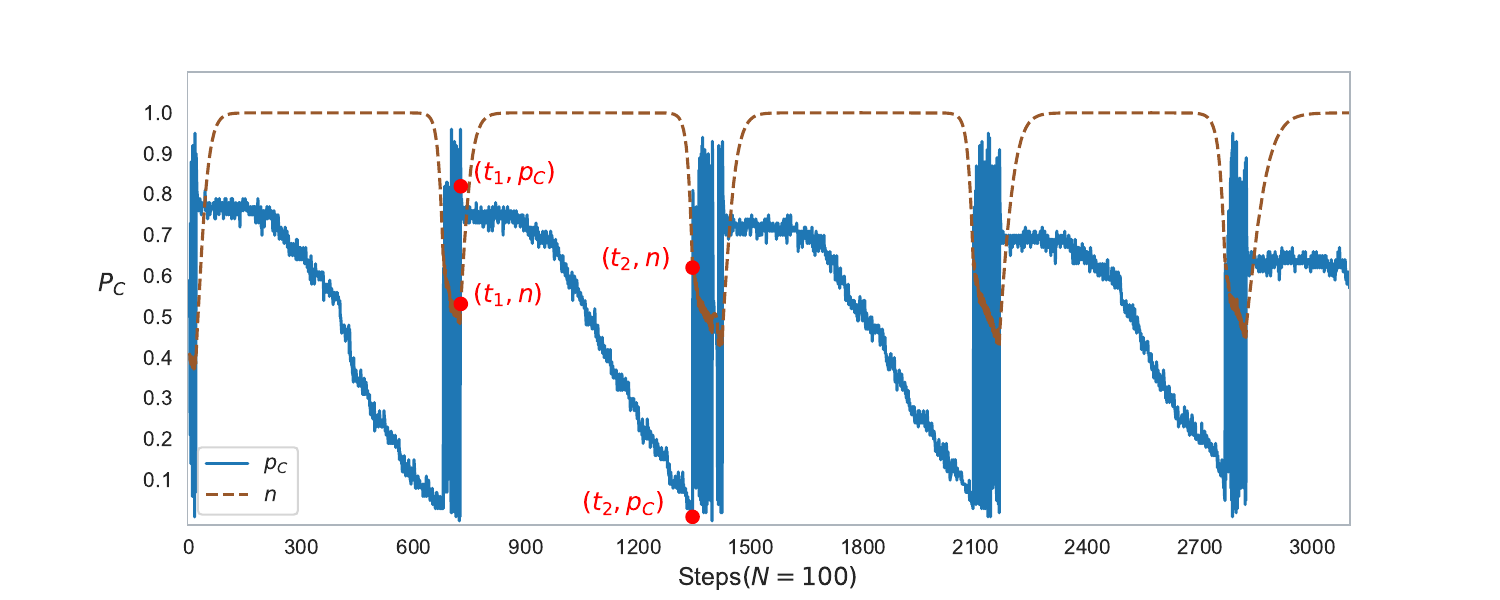}\label{osc in RDE}} 
   
    \subfigure[Cellular diagram during $t=685\sim 760$]{\includegraphics[width=0.4\textwidth]{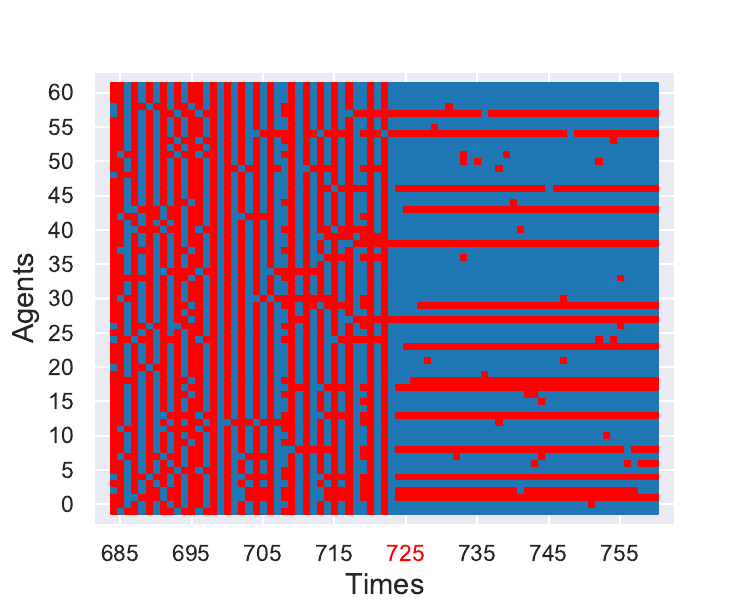}\label{cell1 in RDE}}
    \subfigure[Cellular diagram during $t=1304\sim 1380$]{\includegraphics[width=0.4\textwidth]{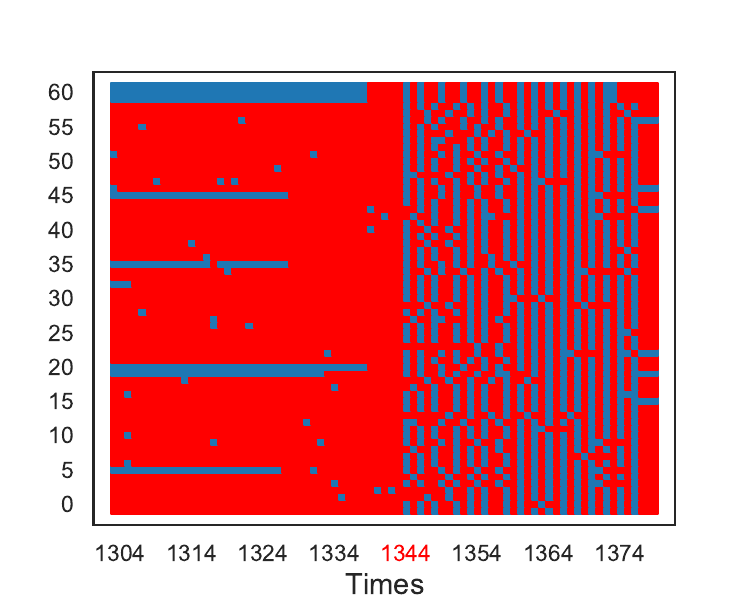}\label{cell2 in RDE}}
        
    \caption{Dynamics of the macro system state in the resource-dependent environment with parameters $N=100,l=0.2,\epsilon=0.04,k_1=0.2$. (a)The system state exhibits periodic oscillations, with $t_1=725$ and $t_2=1344$ marking the critical times for the formation and collapse of the ordered state in the second period. (b)(c)The cellular diagram visually represents the actions of the first 60 agents over time. The state of the group changes from disorder to order at the time $t_1$. Subsequently, agents start a "gold rush" by switching to defectors one by one. At $t_2$ almost all agents become defectors.}
    \label{fig:Oscillations in RDE}
 \end{figure*}
 
\begin{figure}[!t]
    \centering
    \includegraphics[width=\textwidth]{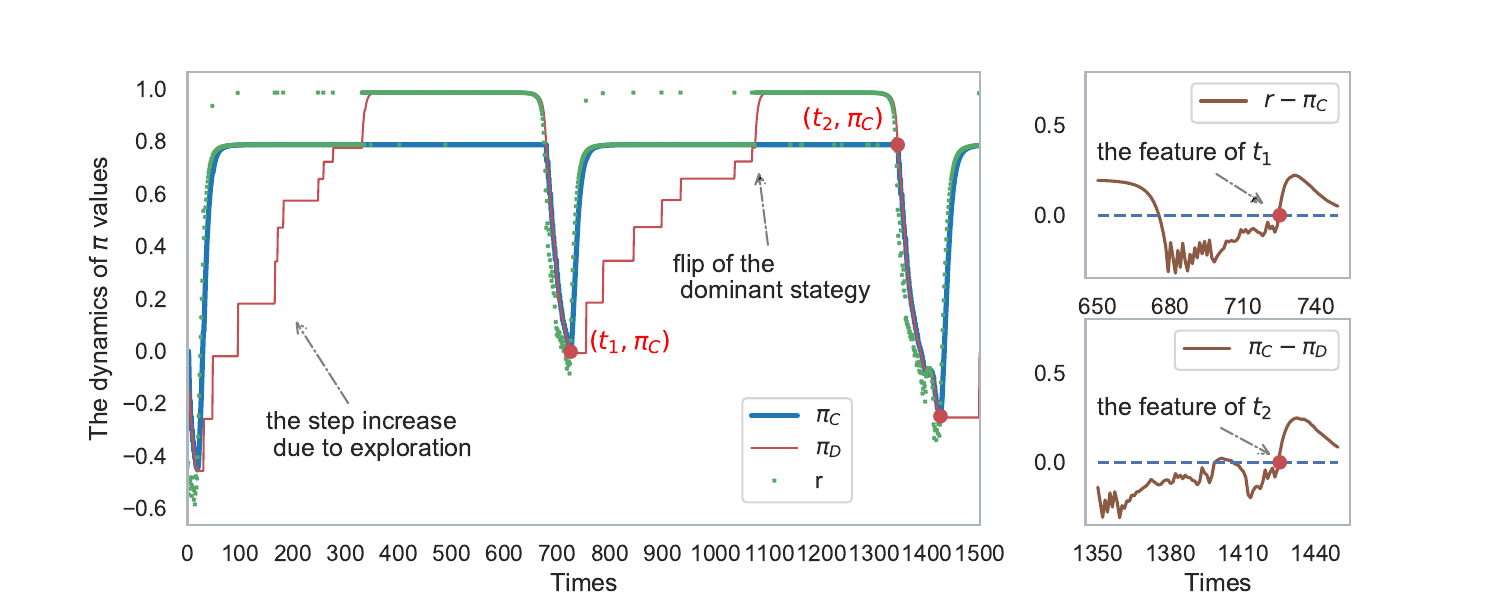}
    \caption{The dynamic $\pi$ value and immediate rewards (indicated by green dots) of Agent11 over the period $t=0$ to $1500$ are shown. In particular, $t_1$ serves as a turning point for the environment. Before $t_1$, Agent11's immediate reward is lower than $\pi_C$ and $\pi_D$. After this point, however, the situation is reversed and Agent11's dominant and dominated strategies are differentiated by the reinforcement of positive rewards, as depicted in the upper right subgraph. At $t=1050$, $\pi_D$ catches up with $\pi_C$, leading to a reversal in Agent11's action.}
    \label{fig:A11}
\end{figure}

\subsection{Formation of a positive feedback loop }

\begin{figure}[!tph]
    \centering
    \includegraphics[width=0.7\textwidth]{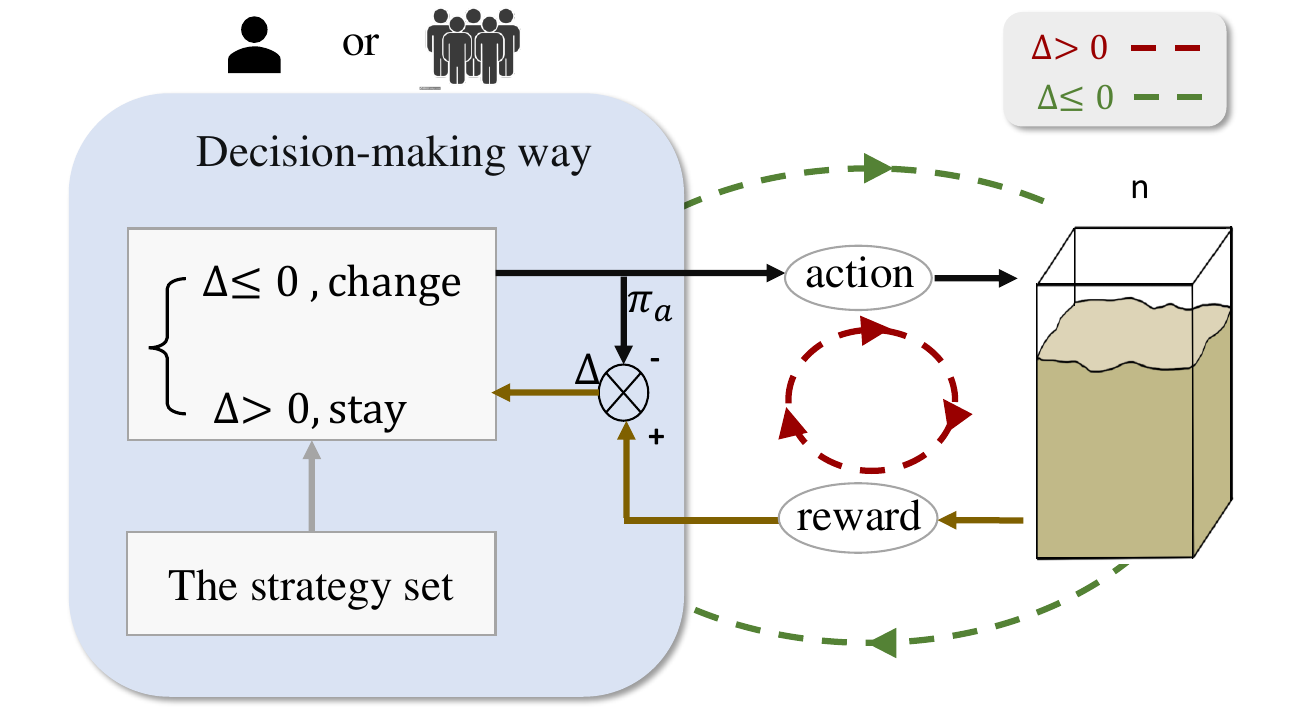}
    \caption{Similarity between the individual and the group behavior. They both behave in a win-stay-lose-change way. When $\Delta=r-\pi_a\le 0$,  the agents or the group act along the trajectory indicated by the green arrow, changing their actions constantly. Until $\Delta>0$ when a positive feedback loop emerges, they take the same action and follow the red circle. Furthermore, the factors contributing to the stability of both the agent's action and the group state are consistent. For the agent, a positive feedback loop is established among the environment, her reward, and her action. Similarly, for the group, the interrelated factors are the environment, the group reward, and the group state.}
    \label{fig:similarity}
\end{figure}

The resource-dependent environment is continuous and there is a significant time lag between the environmental state and group behavior. Despite that, the macroscopic performance still shows a higher frequency of oscillatory patterns. The simulation results of GCM in a resource-dependent environment are shown in Fig.\ref{fig:Oscillations in RDE}. It is noteworthy that the agents perceive the trend of the environment, rather than its state, through rewards. This may not be apparent in a state-dependent environment with discrete states. That is, agents have the ability to detect inflection points of $n$, even without any knowledge of the macro-environment.

Fig.\ref{fig:A11} illustrates this phenomenon, where $t_1$ signifies the inflection point of the environment during the second period, as the environment begins to grow. Prior to $t_1$, the immediate reward $r_i(t)$ of agent $i$ falls below her expected payoffs either $\pi_C$ or $\pi_D$, as depicted in the upper right subgraph of Fig.\ref{fig:A11}. Therefore, regardless of the agent's actions, the expected payoffs continue to decline, resulting in alternating changes in their actions. However, at $t_1$, the agents' immediate rewards exceed their expected payoffs, leading to positive reinforcement of their current actions. Agents experience an upswing in immediate rewards, and therefore their current actions are stabilized as dominant strategies, prompting the group to transition from disordered exploration to a stable and ordered state.

As each agent intermittently adopts an exploratory strategy with a low probability at each step, even the expected payoff of the dominated strategy gradually increases in a favorable environment. Eventually, the cooperators find that the expected payoff of defection outweighs that of cooperation. Consequently, the group start the ``gold rush"\cite{tilman2020evolutionary} by abusing the commons. As more and more cooperators become defectors, the environment deteriorates and agents' rewards fall sharply.The expected value of the dominant strategy, i.e. $D$, then declines until it equals that of strategy $C$ at $t_2$. As a result, the agents and the group enter a new state of disorder.

From the above analysis, we perceive a coherence between the behavior of individuals and groups under both discrete and continuous environments shown in Fig.\ref{fig:similarity}. On the one hand, they both follow the ``win-stay-lose-change" way. Prior to the environmental inflection point, agents alternate between actions $C$ and $D$ and thus the group undergoes continuous state changes from a macroscopic perspective. Nevertheless, once the inflection point is reached, agents experience an upswing in immediate rewards, thereby resulting in positive reinforcement from the environment and the ensuing establishment of dominant strategies. This leads to the emergence of order within the group. On the other hand, the fundamental determinant of which strategy an agent chooses as dominant, and which state of order emerges from group disorder, can be traced back to the formation of a mutually reinforcing positive feedback loop among the environment, immediate rewards, and the individual action(or the group state). In essence, the action or group state that the environment reinforces at the inflection point eventually achieves stability.

\section{Discussion}

In this paper, we set up a multi-agent system called the GCM to study the coupled dynamics between strategies and the macro environment. We simulate two different environments: the state-dependent environment and the resource-dependent environment. Based on the above analysis, we draw the following conclusions.

Fist of all, the coupling effect between macro environment and individual behavior is the key factor to solve the social dilemma. When we analyze the decision-making process of individuals under static scenarios, the optimal action must be defection which leads to the tragedy of commons or the social dilemma due to the higher payoff of defection compared to cooperation. However, when the macro environment are dynamic and coupled with individuals' actions, it is observed that the fitness of different strategies is not fixed, but changeable. The higher payoff of defection can be diluted over time due to the environmental degradation, while cooperation may become the dominant strategy if positively reinforced by environmental feedback. The incentives of different strategies are not constant under changeable environments, as demonstrated in several microbial cases \cite{tilman2020evolutionary}. That is, different strategies constantly competing for survival at the micro level and that one with the highest expected payoff(fitness) will eventually prevail. Cooperation is rightly the result of individual pursuit of self-interest. 

It is worth noting that in the evolutionary process, the group cooperation ratio is not $100\%$; there are always some free-riders. That's because agents have a certain degree of randomness in choosing to be a cooperator or a defector, with the environment being the key determinant. At the inflection point of the environment, the strategy reinforced by the environment finally becomes the dominant strategy for individuals.

As the environment improves, agents are increasingly rewarded as defectors through random exploration. This induces more and more cooperators to switch their actions, eventually leading to the collapse of group cooperation. The self-interest and exploratory tendencies of individuals cause the group state to oscillate between order and disorder.

Secondly, the formation of a positive feedback loop among the environment, immediate rewards, and individual actions (or group states) is the direct cause for the emergence of group cooperation. Insights from simulations demonstrate that, despite agents lacking awareness of the macro-level milieu, they astutely recognize the inflection point solely through immediate rewards. Prior to the inflection point, agents' instantaneous rewards fall short of the expected payoffs for each strategy, compelling them to constantly change their actions. However, when the inflection point is reached, agents' rewards surpass the expected payoff of the current action, triggering a positive feedback loop among the environment, individuals' immediate rewards, and their actions. Agents then experience an upswing in immediate rewards, leading to the establishment of dominant strategies and the emergence of order within the group.

Finally, we find that although individuals adopt an epsilon-greedy algorithm to make decisions, both individuals and groups behave following the ``win-stay-lose-change" pattern. When an individual's immediate rewards are smaller than their expected rewards, they continue changing their actions. Meanwhile, the group's rewards exhibit a downward trend, and its state is subject to constant fluctuations. However, once the inflection point occurs, the rewards for both the group and the individual rise rapidly, resulting in stabilized individual actions and the emergence of group order.

Our article provides a new perspective for understanding the emergence of group cooperation, offering a fresh explanation and revealing the deeper mechanisms that involve individuals and the environment. Currently, humanity faces shared global crises, including overpopulation, air pollution, and nuclear crises, which require joint efforts from all countries. Our study provides a new approach to solving these problems. As countries become increasingly interconnected in every aspect, none can remain immune to the macro environment formed by the collective actions of all. As long as interests align, cooperation becomes natural and inevitable.

\section*{Acknowledgements}

We thank H.Zhao, D.Zhang and all the participants for their valuable feedback and constructive suggestions on this research.

\bibliographystyle{elsarticle-num} 

\appendix 

\section*{Supplement Information} 
\addcontentsline{toc}{section}{Supplement Information} 


\section{Methods}

In fact, the Group Coupling Model(GCM) refers to experiments on the cognitive behavior of animals, i.e. the pigeons' shape-color recognition experiments. In these experiments, pigeons strive to obtain more rewards by learning the association between lateral strokes and food. Remarkably, the utilization of Q-learning has been demonstrated to replicate the learning processes of pigeons \cite{rose2009theory,glascher2010states,ruan2009skinner}. Consequently, an agent is introduced to replace the pigeons in the coupled model, acting the same way that maximizes her payoff through continuous learning of the relationship between strategies and rewards. The distinction between these two lies in the environmental conditions: the animal cognitive experiment involves a static environment typically with a single individual, whereas the environment in GCM is dynamic and coupled with the group's actions which is a multi-agent system."

\begin{figure*}[!b]
    \centering
    \subfigure[]{\includegraphics[width=0.5\linewidth]{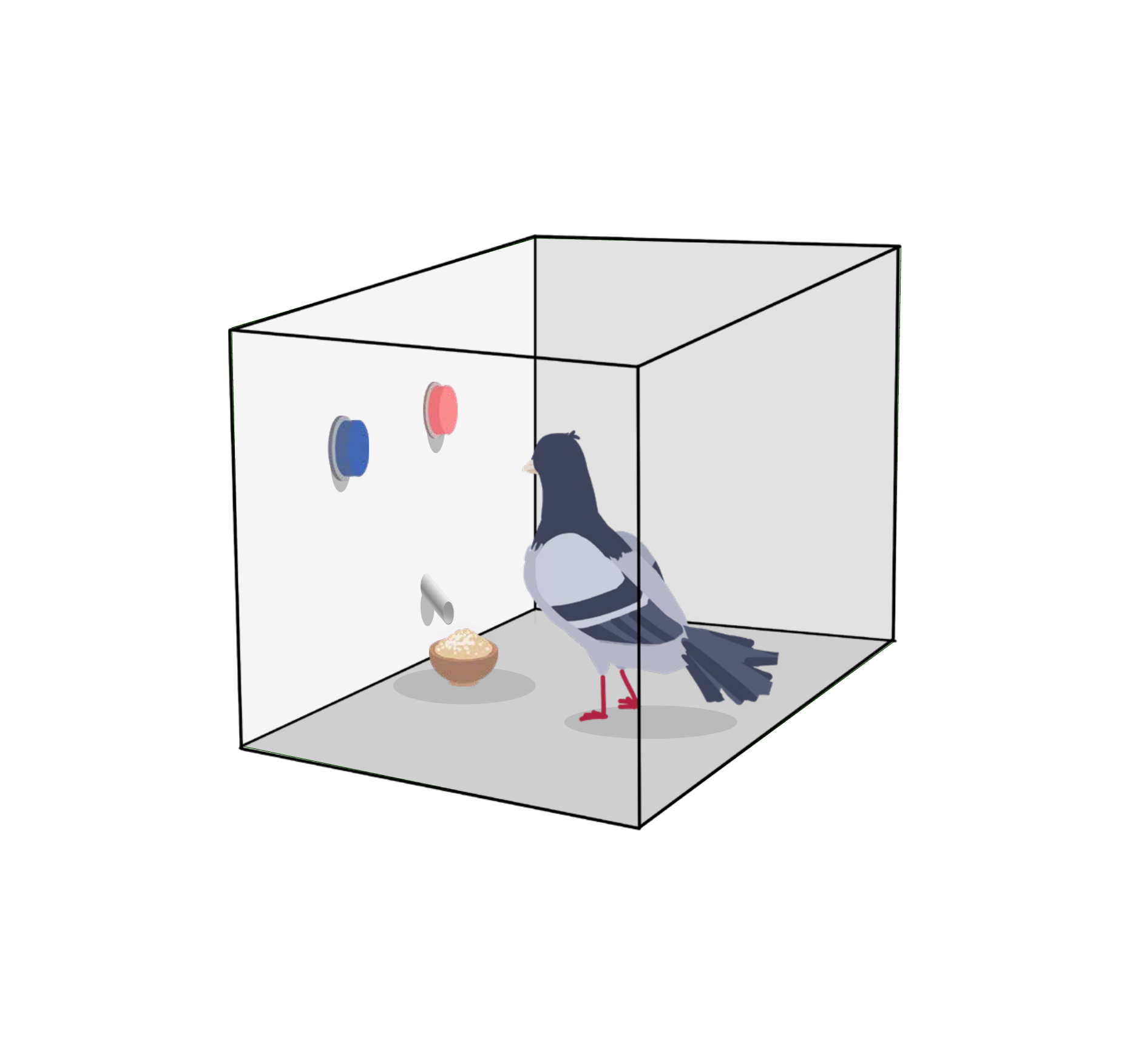}\label{one_p}} 
    \subfigure[]{\includegraphics[width=0.4\linewidth]{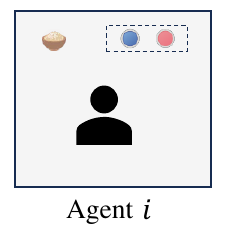}\label{one_a}} 
   
    \caption{(a)Pigeons' shape-color recognition experiments; (b) The simulation configuration for an agent in GCM. }
    \label{si:pigeon}
\end{figure*}

\subsection{Simulating and differentiating different strategies in GCM}\label{analysis}

The GCM is a homogeneous system with no spatial structure, where each agent faces the same experimental configuration, i.e. two different strategy choices and the resulting payoffs in Fig.\ref{si:pigeon}. So how do agents differentiate between different strategies? In the pigeons' shape-color recognition experiments, pigeons discriminate between different strategies not based on the color and shape of the buttons, but rather on the rewards obtained upon pressing the buttons. As a result, we can simulate different strategies($C$ and $D$) by associating them with different rewards. That is, simulations of different strategies can be realized by the designed payoff matrix, as shown in Fig.\ref{fig:model}.

\subsection{Comparison between self-learning method and the replication equation method}

 Typically, research on evolutionary games with environmental feedback predominantly employs replicator equations to derive the evolutionary dynamics from the group perspective, in which

\begin{equation}\label{replicator}
    \dot{x}=x(1-x)\left[R_C(n)-R_D(n)]\right. 
\end{equation}
 where $R_C(n)$ and $R_D(n)$ denote the expected payoff of cooperators and defectors, respectively. $x$ refers to the proportion of cooperators in the population.

We incorporate the payoff matrix in GCM into the Eq.\ref{replicator}, and assume that the payoffs from an individual are linear in both the frequency of cooperators in the population $x$ and the environment $n$. That is,

\begin{equation}\label{eq7}
\begin{aligned}
    \left[R_C(n)\quad R_D(n)\right]=&(1-n) *\left[\begin{array}{ll}-1& -1\end{array}\right]\\
                    &+n *\left[\begin{array}{ll}1-c& 1\end{array}\right]
\end{aligned}
\end{equation}
Thus,
\begin{equation}\label{eq8}
    \left\{\begin{matrix}
\begin{aligned}
R_C(n)=&2*n-1-n*c \\
R_D(n)=&2*n-1 
\end{aligned}
\end{matrix}\right.
\end{equation}

Following the function of the evolution environment\cite{weitz2016oscillating}, we have:

\begin{equation}\label{envrionment}
\begin{aligned}
    \dot{x}=& -x(1-x)*n*c \\ 
     \dot{n}=& kn(1-n)\left[(1+\theta)x-1]\right. 
\end{aligned}
\end{equation}

There are three kind of fixed points of GCM by replicator dynamics with feedback-evolving games. That is, $(i)$ $(x, n^*=0)$, any $x$ in a degraded environment; $(ii)$ $(x^*=0, n^*=1)$, defectors in a replete envrionment; $(iii)$ $(x^*=1, n^*=1)$, cooperators in a replete envrionment.

The stability of the solution can be judged by calculating the eigenvalues of the Jacobian matrix of the system. The rules for stability are as follows:
\begin{enumerate}
    \item If the real part of all the eigenvalues of the Jacobian matrix is negative, then the fixed point is stable (or attractive).
    \item If the real part of any eigenvalue of the Jacobian matrix is positive, then the fixed point is unstable.
    \item If the real part of the eigenvalue is 0, then we cannot be sure of the stability of the fixed point and a deeper analysis is needed.
\end{enumerate}

The Jacobian matrixs of the three kind of fixed point are:

\begin{equation}
\begin{aligned}
    &J(0, 0)=  \left[\begin{matrix}0 & - c x \left(1 - x\right)\\0 & k \left(x \left(\theta + 1\right) - 1\right)\end{matrix}\right] \\
    &J(0, 1)=  \left[\begin{matrix}- c & 0\\0 & k\end{matrix}\right] \\
    &J(1, 1)=  \left[\begin{matrix}c & 0\\0 & - k \theta\end{matrix}\right]
\end{aligned}
\end{equation}

\begin{figure*}[!tph]
    \centering
    \subfigure[The numerical result by replicators] {\includegraphics[width=0.7\textwidth]{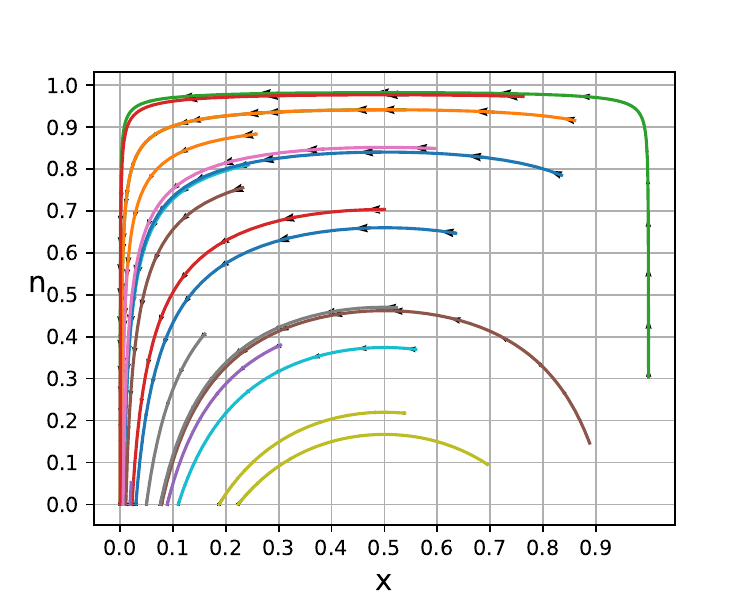}\label{pp1}} 
    \subfigure[The simulation result by self-learning]{\includegraphics[width=0.7\textwidth]{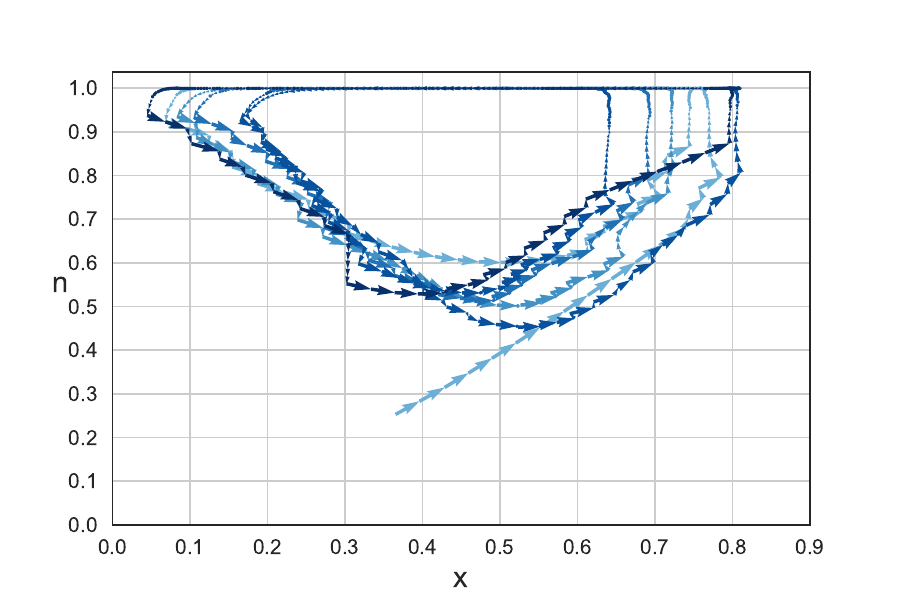}\label{pp2}} 
   
    \caption{Different phase plane dynamics of x-n between the replicators and self-learning method with $k=0.2$. They share the same payoff matrix.(a) Different colors represent trajectories with different initial conditions. (b) The initial condition is $n=0.4$. The blue color shows the evolution of time from light to dark.  }
    \label{si:phase_plane}
\end{figure*}

The analysis for stability is as follows:
\begin{enumerate}
    \item In $(x, 0)$, Jacobian matrix characteristic value of $\{k *(\theta x + x - 1):1, 0, 1\}$. One of these two eigenvalues is 0, and the other may be positive or negative, depending on the specific values of $\theta$ and $x$. Thus, this fixed point may be unstable if $k*(\theta x + x-1) > 0$ or stable if $k*(\theta x + x-1) < 0$. It may also be neutral stable if $k*(\theta x + x-1) = 0$. Therefore, we cannot directly determine the stability of this point and need specific parameter values.

    \item At the point $(0,1)$, the eigenvalue of the Jacobian matrix is $\{-c: 1, k: 1\}$. Where $-c$ must be negative and $k$ must be positive. A positive eigenvalue indicates that the point is unstable in a certain direction. Therefore, we can conclude that the point $(0,1)$is unstable.
    
    \item At the point $(1,1)$, the eigenvalue of the Jacobian matrix is $\{c: 1, -k*\theta: 1\}$. The eigenvalue $c$ is positive, indicating that the point is unstable in a certain direction. The eigenvalue $-k*\theta$ may be positive or negative, depending on whether $\theta$ is negative. But in most biological models, we expect $\theta$ to be positive, so this eigenvalue is usually negative. However, since there is a positive eigenvalue, we can still conclude that the point $(1,1)$is unstable.

\end{enumerate}

As a consequence, we will expect an eventual deterioration of the environment by the replicator equations. This result is due to the fact that the replicator equation compares the rewards of different strategies at the same time, in which the payoff of strategy $D$ is definitely more than that of strategy $C$ whenever the environment is replete or depleted. Even if the payoffs are environment-dependent, it is still essentially under a static environment. Due to the similarity in their underlying principles, both the imitation method and the replicator equation are expected to yield identical results.

However, in the case of self-learning methods shown in Eq.\ref{learning}, there is a long-range memory between individual income and the environment, that is,

\begin{equation}\label{time}
        \pi_{ai}(t) =l*[r_{t-1} +(1-l)*r_{t-2}+...+(1-l)^{t-1}r_{0}] 
    \end{equation}
when initial value of $\pi_{ai}(0)$ is $0$. As a result, the expected payoffs associated with different strategies exhibit variability over time. The immediate higher rewards of defection are susceptible to dilution over time due to the degradation of the macro environment.  Similarly, the relatively modest rewards of cooperation may undergo amplification as a consequence of enhanced environment. A multitude of scholarly sources have revealed that the kind of  ``trial and error'' potentially represents a more pervasive and ubiquitous paradigm, capable of elucidating the behavioral dynamics observed in diverse biological, microbial, and botanical entities\cite{gagliano2014experience,burton2021payoff,trewavas2003aspects}.

In addition, agent-based models adopt a different paradigm, envisioning a world in which decision-making is decentralized \cite{macy2002learning}. This approach allows for individual agents to be observed as objects of study\cite{zhang2020oscillatory,geng2022reinforcement}, facilitating comparative analysis of the emergence and evolution of cooperation from both macro(group) and micro(individual) perspectives.

\section{Extra Figures}

\begin{figure}[!t]
    \centering 
    \subfigure[]{\includegraphics[width=0.48\textwidth]{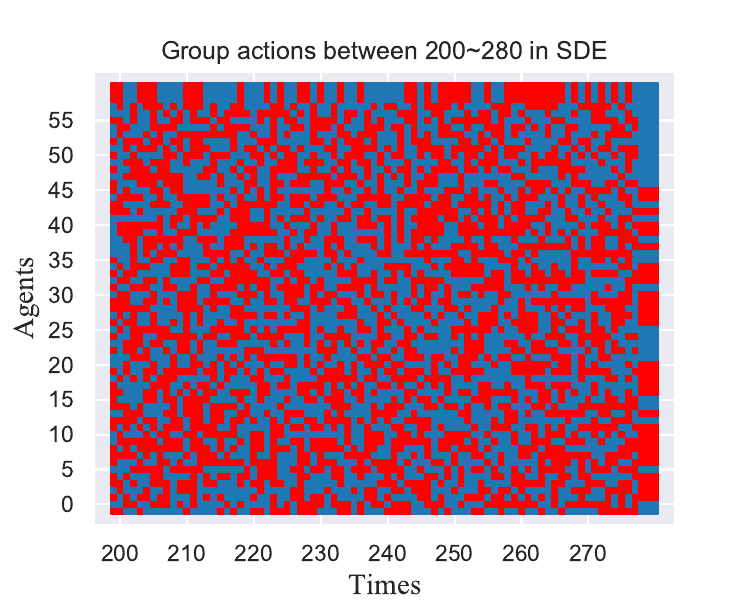}\label{fig:s1}}
    \subfigure[]{\includegraphics[width=0.48\textwidth]{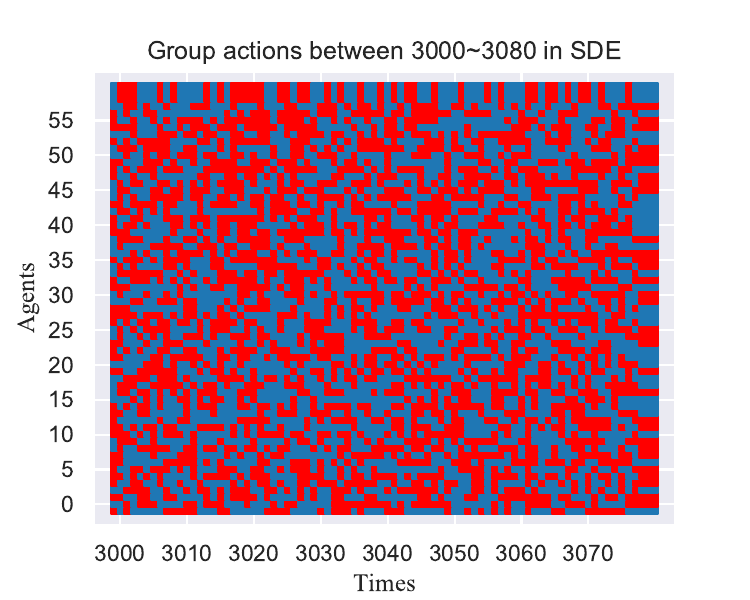}\label{fig:s2}}
    \subfigure[]{\includegraphics[width=0.48\textwidth]{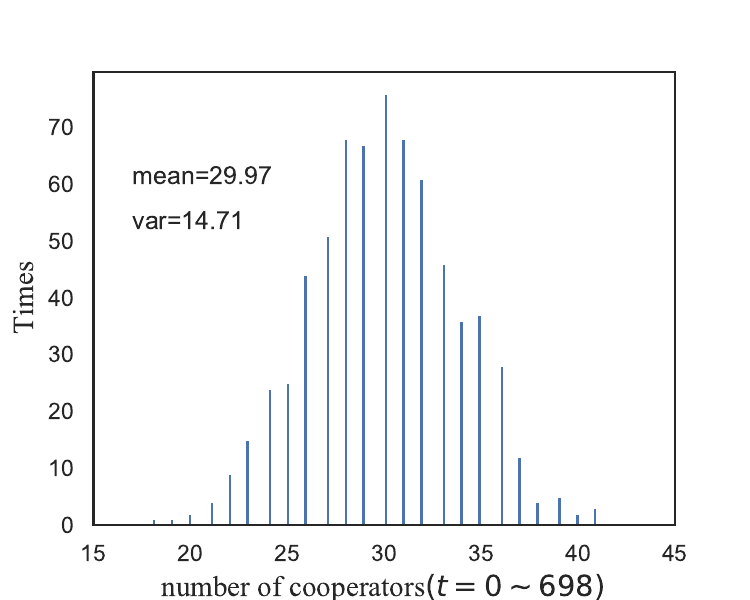}\label{fig:s3}}  
     \subfigure[]{\includegraphics[width=0.48\textwidth]{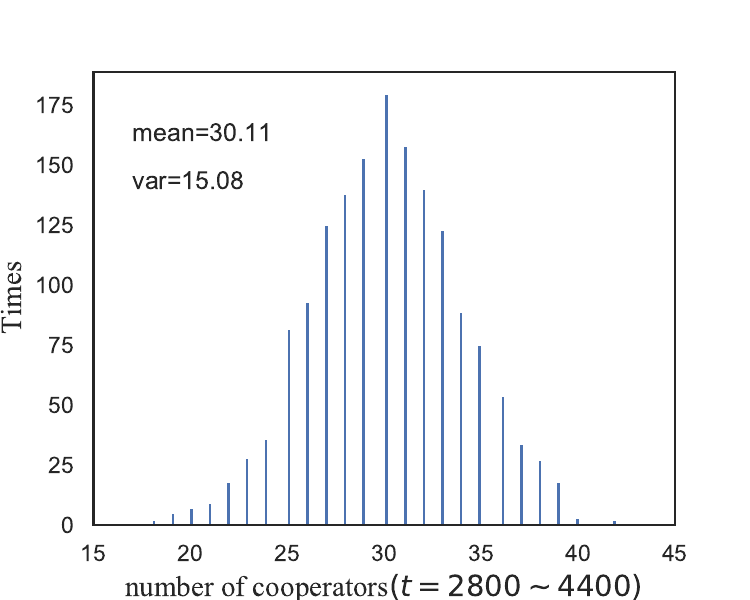}\label{fig:s4}}
        
    \caption{Search patterns of the group in the state-dependent environment (SDE). (a)(b) The cellular diagrams of the group in SDE at $t=200\sim 280$ and $t=3000\sim 3080$, respectively. (c) (d) The group cooperation rate shows a normal distribution during this period, indicating that each individual is conducting a random search between strategies $C$ and $D$. This behavior is a result of both agents having $\pi_C$ and $\pi_D$ values of $-1$, and agents employing the $\epsilon$-greedy algorithm being required to choose actions randomly.}
    \label{fig:normal_sde}
\end{figure}

\begin{figure}[!t]
    \centering 
   
    \subfigure[]{\includegraphics[width=0.48\textwidth]{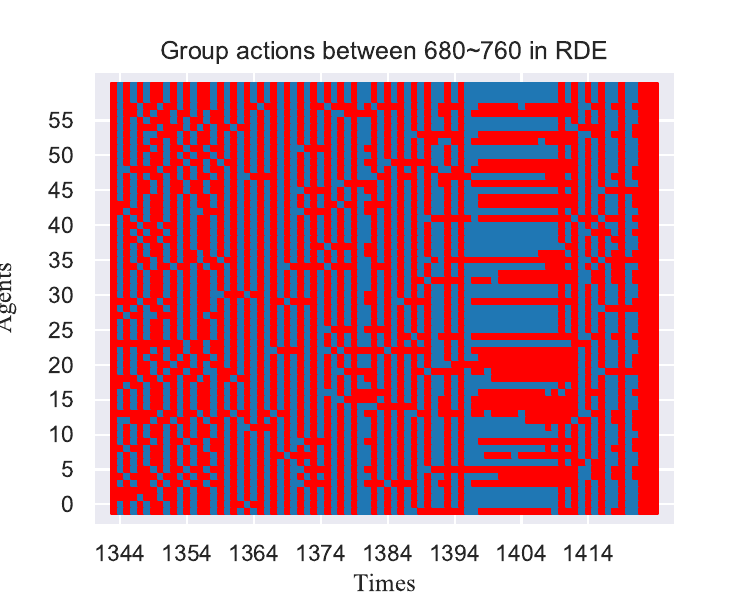}\label{fig:s5}}
    \subfigure[]{\includegraphics[width=0.48\textwidth]{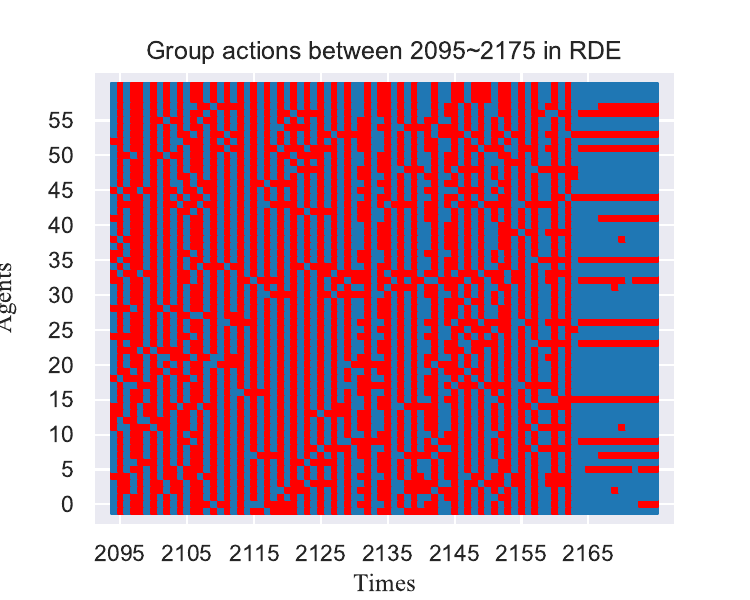}\label{fig:s6}}
   
    \subfigure[]{\includegraphics[width=0.48\textwidth]{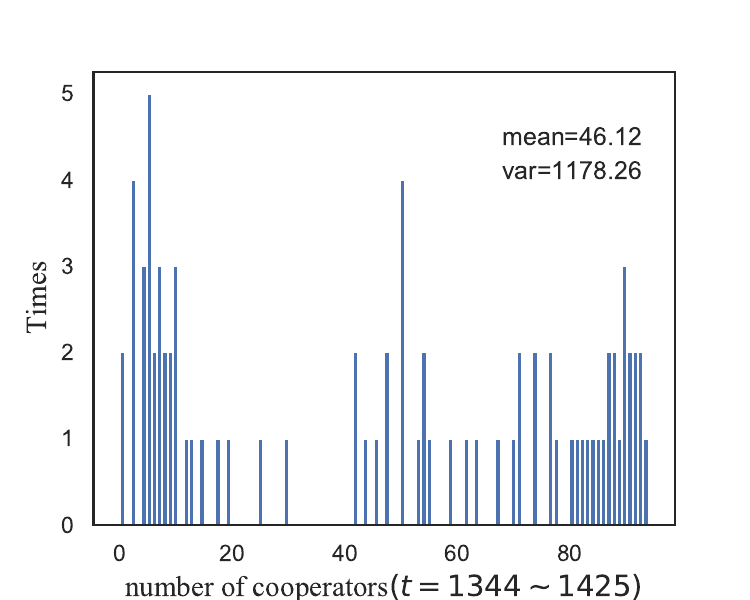}\label{fig:s7}}
    \subfigure[]{\includegraphics[width=0.48\textwidth]{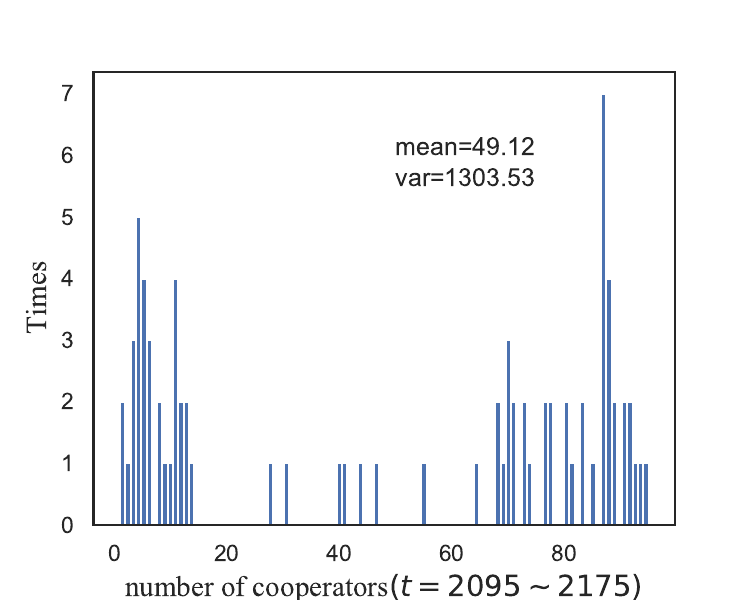}\label{fig:s8}}
        
    \caption{Search patterns of the group in the resource-dependent environment (RDE). (a)(b) The cellular diagrams of the group in RDE at $t=1344\sim 1424$ and $t=2095\sim 2175$, respectively. (c)(d) The group cooperation rate shows a relatively uniform distribution during those period, which is very different from that in the SDE. That's because $\pi_C$ and $\pi_D$ of agents alternately decrease, so that the group cooperation rate is distributed at both ends of the district.}
    \label{fig:normal_rde}
\end{figure}

\begin{figure}[!t]
    \centering 
    \includegraphics[width=\textwidth]{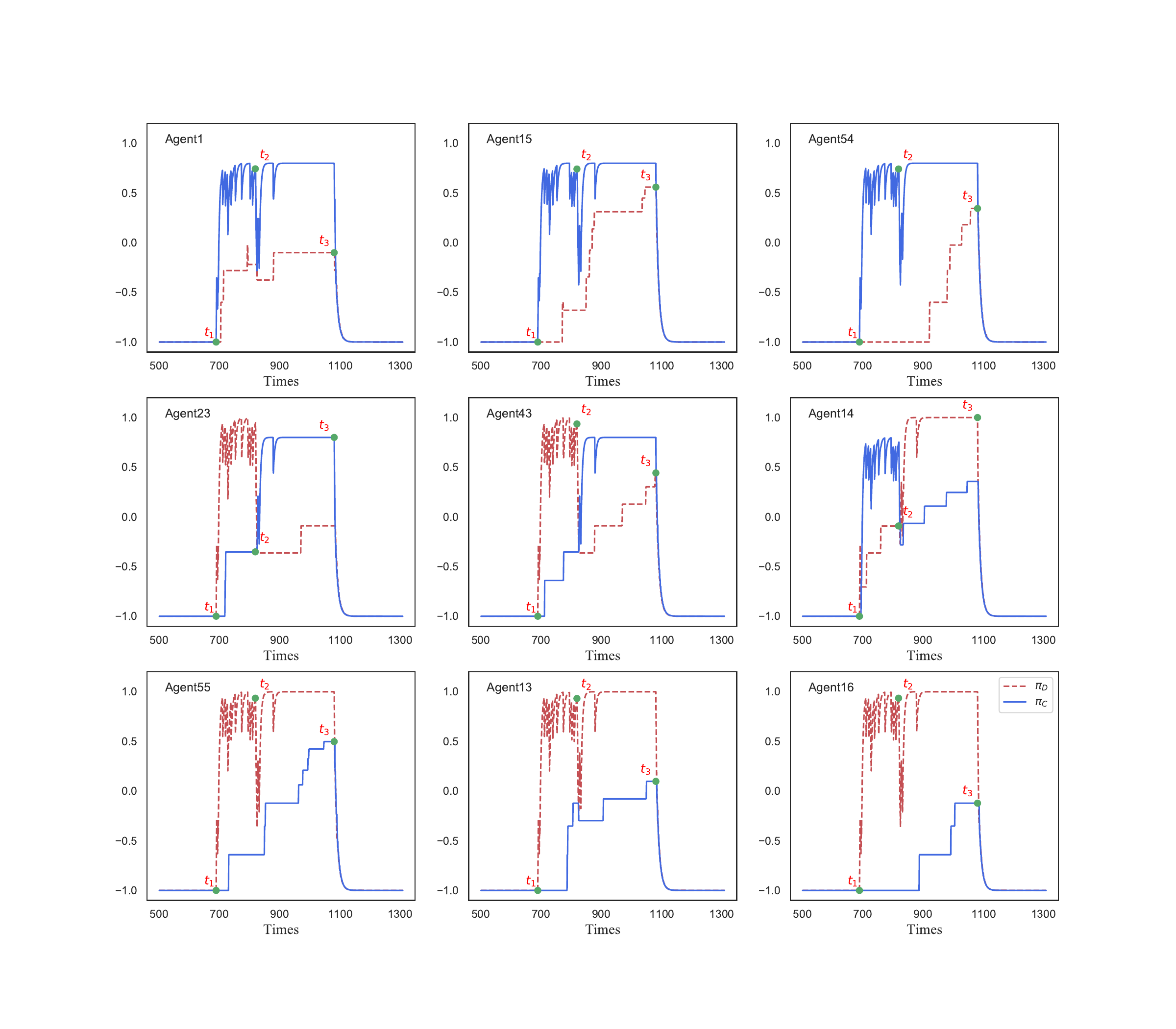}
    \caption{The dynamics of the $\pi$ value of different agents in SDE. Although each agent uses the same update method and decision-making approach, the $\pi$ value of them evolves in a completely different way. As shown in the figure, the three agents in the first row all adopt cooperation as the dominant strategy in the first oscillatory cycle; the three agents in the second row flip their dominant strategies during this time interval, and another three agents in the third row all adopt defection as their dominant strategy. Besides, even though the dominant strategies of agents are the same, the dynamic of the $\pi$ value is completely different in details. The moments $t_1$, $t_2$, and $t_3$ in the figure correspond to the same points in Fig.2(a) of the main text, respectively.}
    \label{fig:agents in SDE}
\end{figure}

\begin{figure}[!t]
    \centering 
    \includegraphics[width=\textwidth]{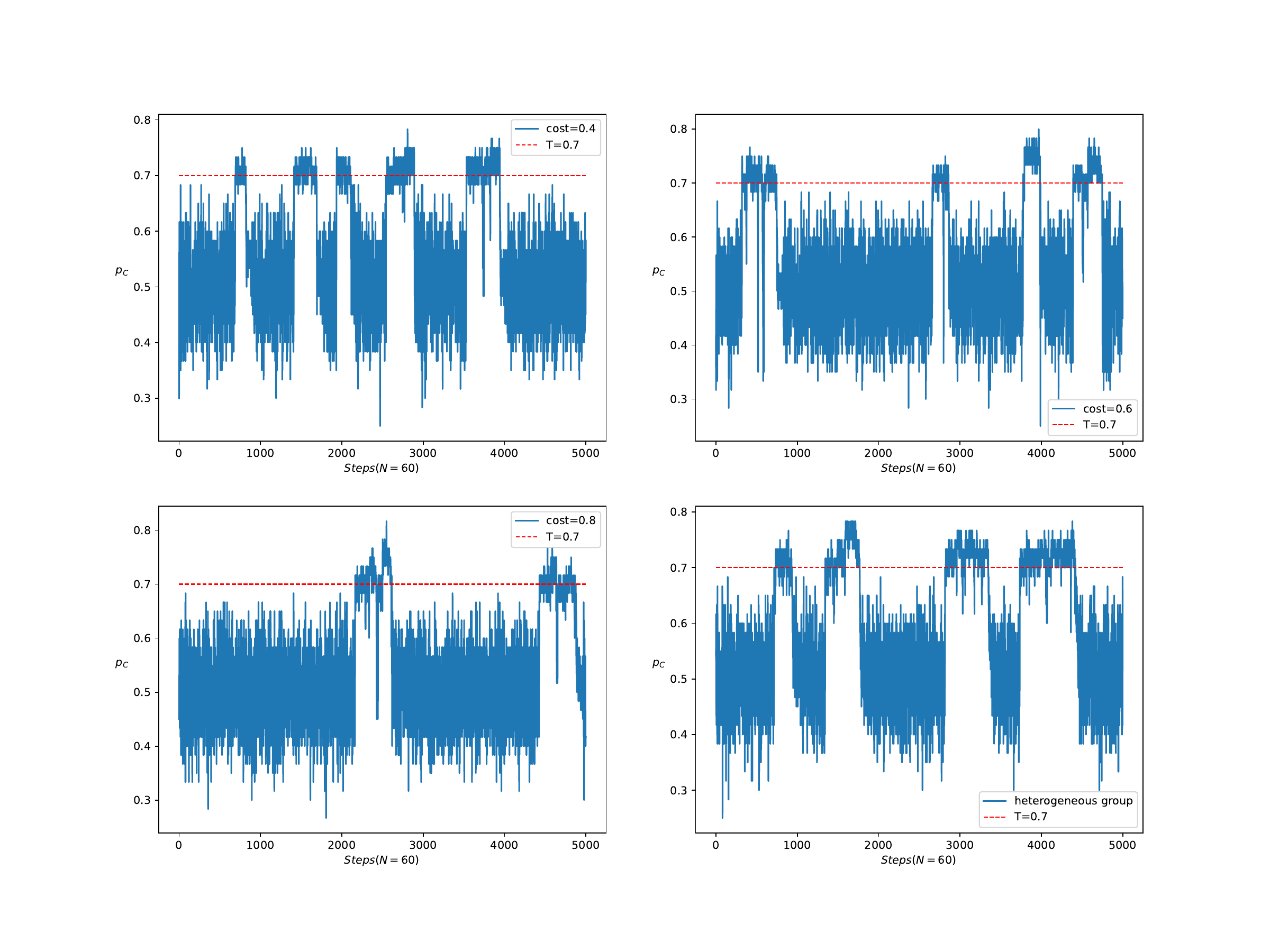}
    \caption{Evolutionary dynamics of group cooperative rate under different parameters. The first three diagram is draw by homogeneous groups, in which the parameters are $c=0.4$, $0.6$, and $0.8$, respectively. The result of oscillation in main text is stable, and not significantly affected by the value of cost. The last one is the evolutionary dynamics of cooperative rate for heterogeneous groups in AIM which indicates the heterogeneity of the group does not significantly affect the oscillation. Here the heterogeneous group includes individuals who use either the '$\epsilon-greedy$' or the 'Boltzman' method for decision-making, and their learning rate and explorative parameters are different from each other. The range of values for the learning parameter $l$ is $0.05\sim 0.3$, and the range of values for $\epsilon$ is $0.01\sim 0.08$.}
    \label{fig:parameter}
\end{figure}

\begin{figure}[!b]
    \centering 
    \subfigure[The relation between $n-K$ and $n$ in RDE]
    {\includegraphics[width=0.7\textwidth]{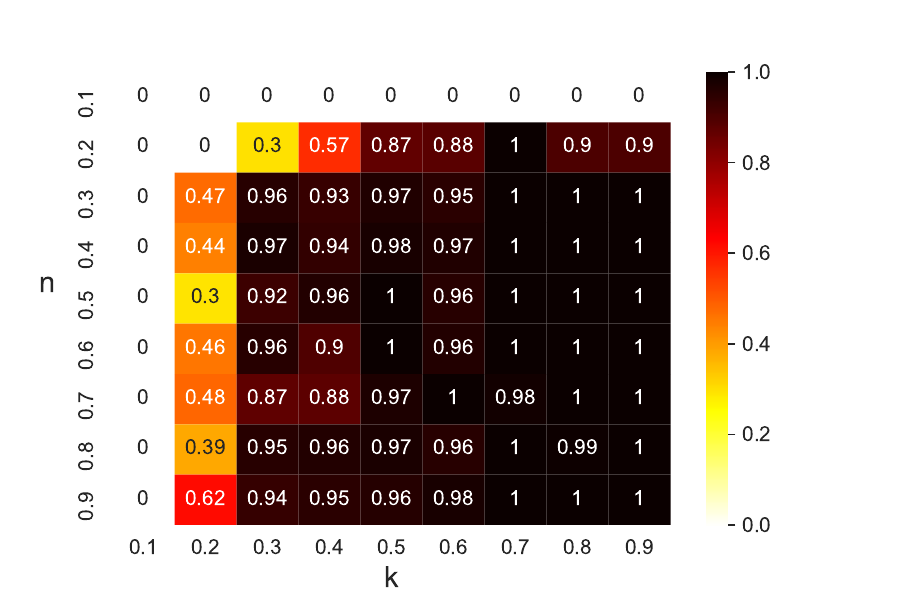}\label{nkn}}
    \subfigure[The relation between $n-K$ and $p_C$ in RDE]
    {\includegraphics[width=0.7\textwidth]{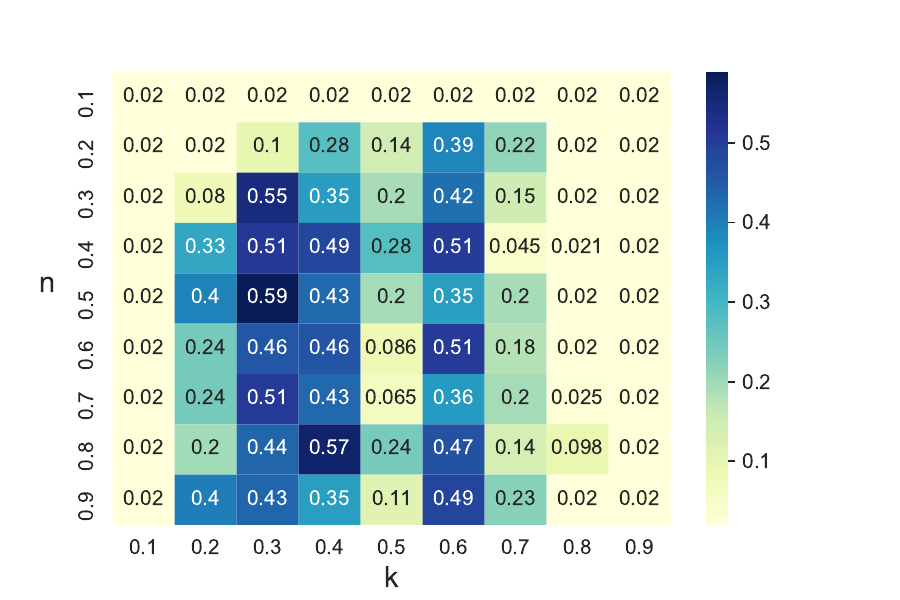}\label{nkp}}
    \caption{The effect of initial environment value $n$ and environmental change rate $k$ on the system state,including the envrionmental state and the cooperative level. From the figure, we can see that the system state is significantly affected by the value of $k$ , when $k<0.1$, the system state is decreased to 0. When $k\geq 0.7$, due to the rapid change of the environment, it quickly reaches 1, while the group cooperation rate tends to 0. The values were obtained by running 20 random variations with the corresponding parameters and averaging the latter 200 time steps. The oscillation of the system state occurs when $ \theta \geqslant 1,n\in [0.2,0.9],k\in [0.2,0.7)$.}
    \label{fig:parameter in RDE}
\end{figure}

\begin{figure}[!t]
    \centering 
    \includegraphics[width=\textwidth]{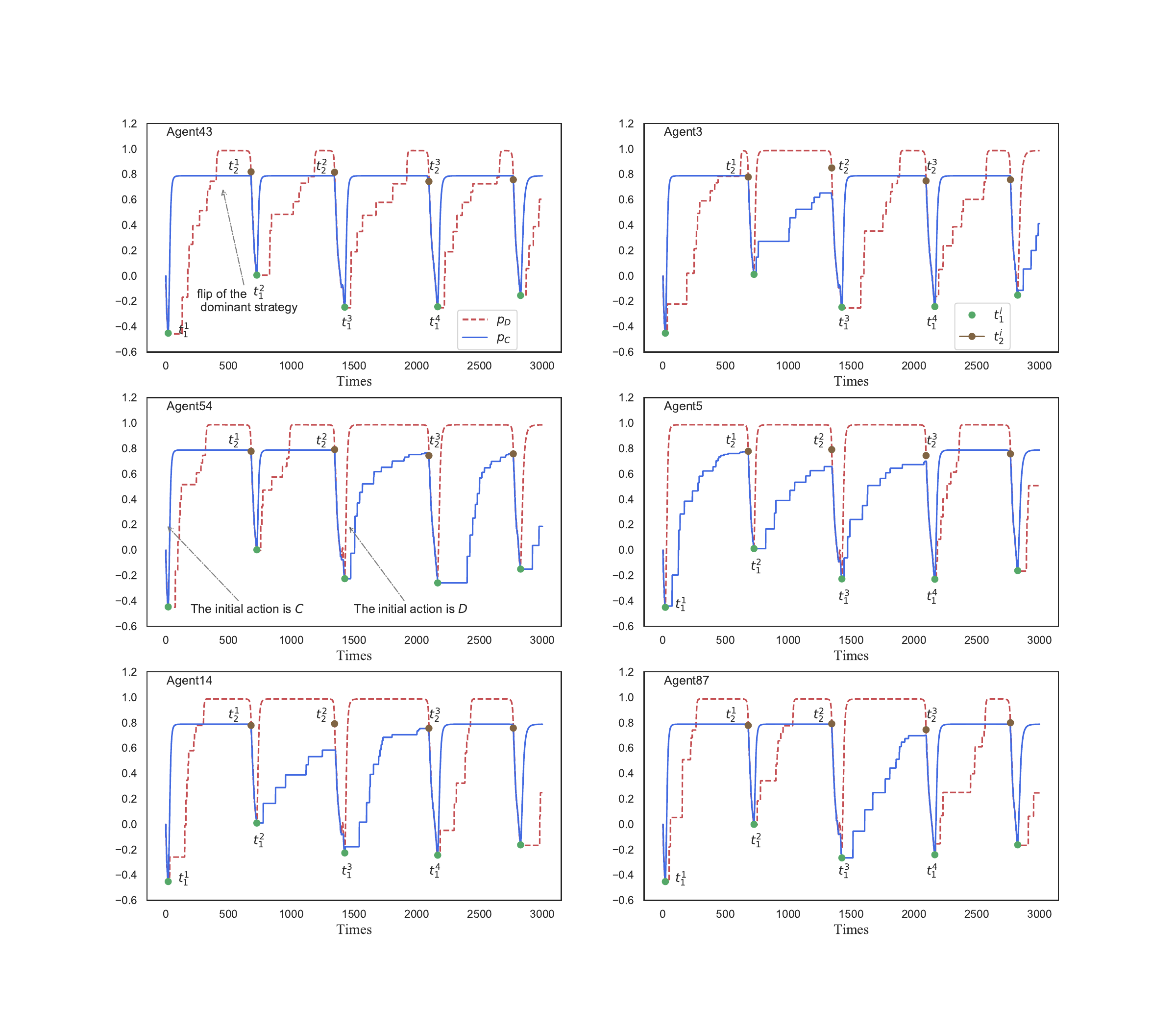}
    \caption{The dynamics of the $\pi$ value for six agents in four oscillation cycles is given in the figure. These agents have different dominant strategies in different oscillation cycles. For example, the dominant strategy of agents in the first row is $CCCC$ and $CDCC$, respectively. And that in the second row is $CCDD$ for Agent54 and $DDDC$ for Agent5. Which strategy the agent chooses as its dominant strategy has some randomness, that is the action the agent takes at the inflection point (the green point in the figure),  becomes its dominant strategy due to the positive feedback from the environment. The brown points in the figure are the end points of the ordered states of the group, denoted by $t_2^i$, and the green points are the start points of those states, denoted by $t_1^i$, with the superscript $i$ denoting the cycle number. And one rise and fall of the $\pi$ value for each agent corresponds to one cycle in Fig.\ref{fig:Oscillations in RDE}\subref{osc in RDE} of the main text.}
    \label{fig:agent in RDE}
\end{figure}

\begin{figure}[!t]
    \centering 
    \subfigure[The dynamic $\pi$ value of Agent11]
    {\includegraphics[width=\textwidth]{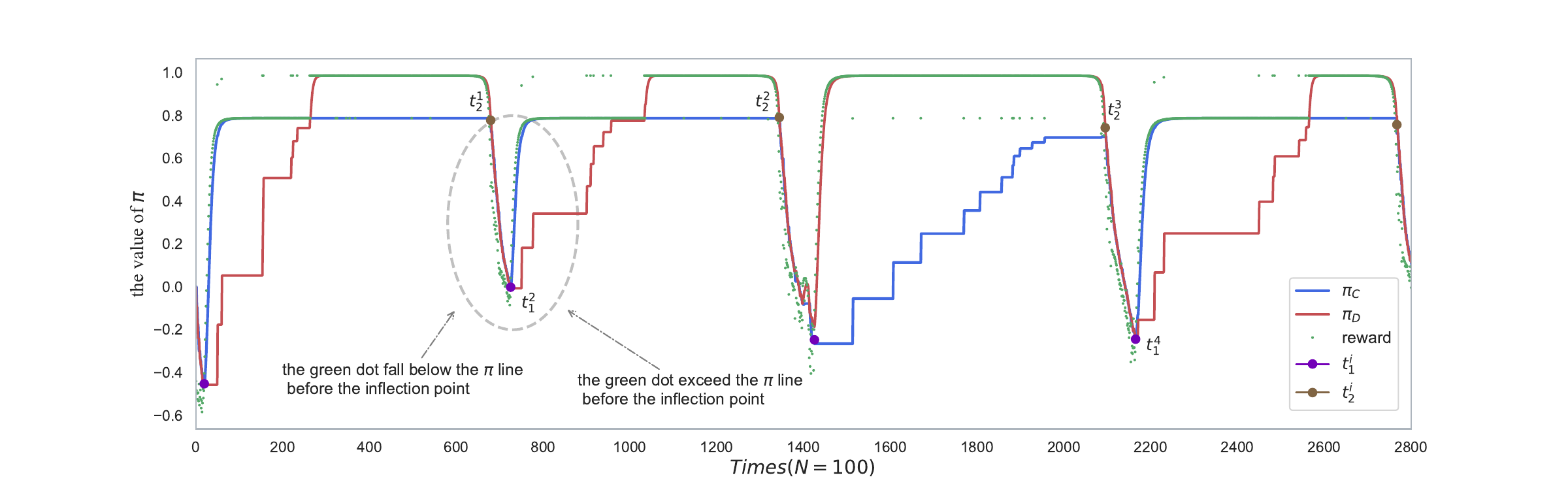}\label{A11Q}}

    \subfigure[The feature of the point $t_2$]
    {\includegraphics[width=0.9\textwidth]{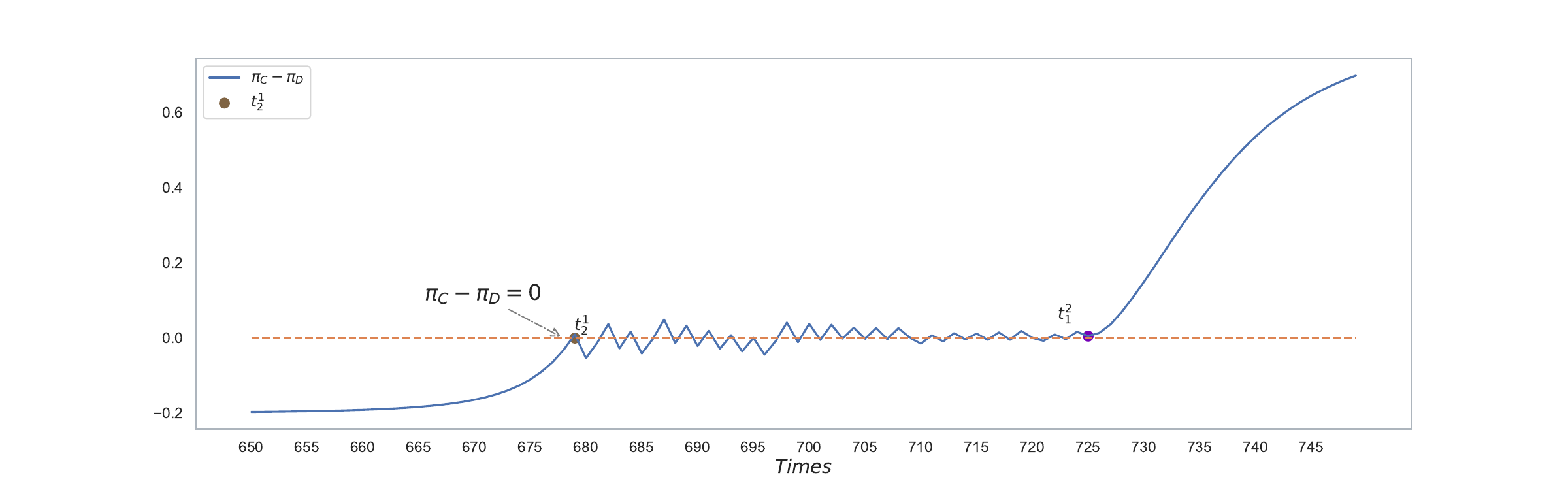}\label{feature1}}
    
    \subfigure[The feature of the point $t_1$]
    {\includegraphics[width=0.9\textwidth]{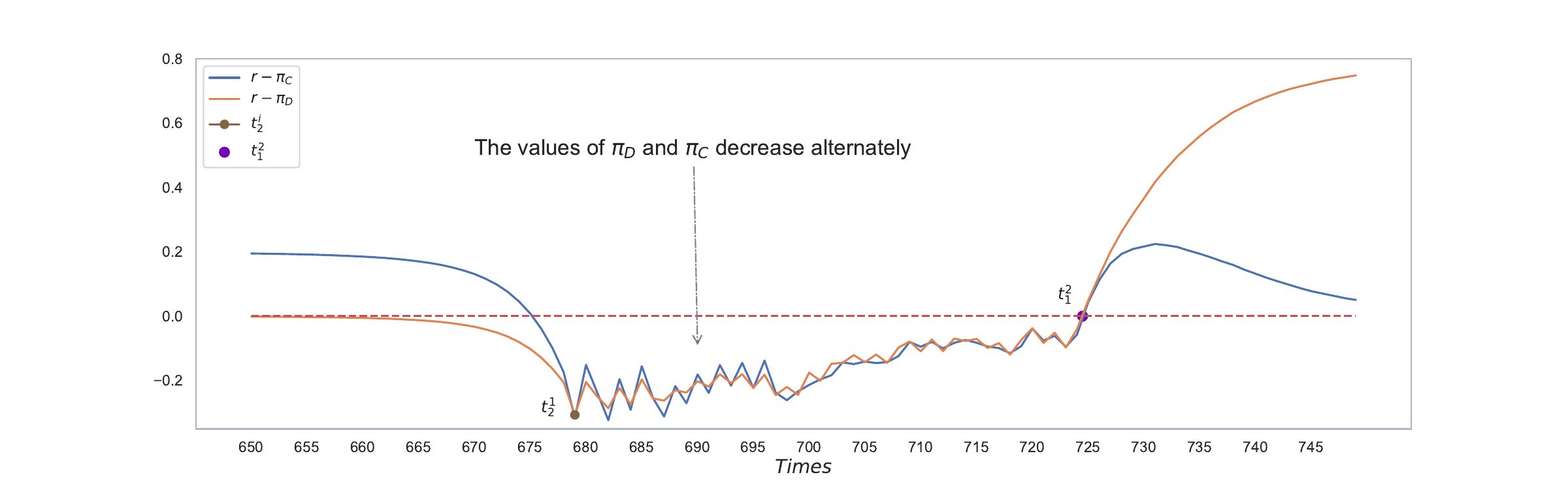}\label{feature2}}
    \caption{These are the enlarged diagrams of Fig.4(d) in the main text, which provides clearer details. (a)We can get that before $t_1^2$, the green dots are below the red and blue curves, indicating that the agent's immediate payoff is less than $\pi_C$ and $\pi_D$. After this point, the green dots are above the those colored curves. And the blue curve rises rapidly, indicating that strategy C becomes the dominant strategy for the agent. (b) The characteristic of the end of the ordered state of the group is $\pi_C=\pi_D$. After this point, as the agent's immediate payoff is less than $\pi_C$ and $\pi_D$, she alternates between using C and D strategies. This is shown as a jagged line in (c). $t_1^2$ is the start of the second ordered state on the time axis, characterized by the agent's immediate payoff being greater than $\pi_C$ and $\pi_D$.}
    \label{fig:feature}
\end{figure}






\end{document}